\documentclass[fleqn,10pt]{wlscirep}
\usepackage[utf8]{inputenc}
\usepackage[T1]{fontenc}
\usepackage{wrapfig}
\usepackage{bm}

\title{\textit{A}V$_3$Sb$_5$ Kagome Superconductors:  Progress and Future Directions}

\author[1,*]{Stephen D. Wilson}
\author[2]{Brenden R. Ortiz}
\affil[1]{Materials Department, University of California Santa Barbara, California 93106, USA}
\affil[2]{Materials Science and Technology Division, Oak Ridge National Laboratory, Oak Ridge, 37831, Tennessee, USA}

\affil[*]{e-mail: stephendwilson@ucsb.edu}

\begin{abstract}
	The recent discovery of the \textit{A}V$_3$Sb$_5$ (\textit{A}=K, Rb, Cs) kagome superconductors launched a growing field of research investigating electronic instabilities in kagome metals. Specifically, the \textit{A}V$_3$Sb$_5$ family naturally exhibits a Fermi level tuned to the Van Hove singularities associated with the saddle points formed from the prototypical kagome band structure. The charge density wave and superconducting states that form within the kagome networks of these compounds exhibit a number of anomalous properties reminiscent of theoretical predictions of exotic states in kagome metals tuned close to their Van Hove fillings.  Here we provide an overview of the key structural and electronic features of \textit{A}V$_3$Sb$_5$ compounds and review the status of investigations into their unconventional electronic phase transitions.
\end{abstract}
\begin{document}
	
	\flushbottom
	\maketitle
	
	\thispagestyle{empty}
	
	\section*{Main text}
	
	\section*{Introduction}
	
	Kagome lattices, networks of corner sharing triangles,\cite{syozi1951statistics} have long been crucial building blocks for a range of unconventional states sought in condensed matter physics.  Insulating compounds built from kagome networks of localized spins provide extremely rich platforms for studying magnetic frustration and potential spin liquid states predicted to form from the geometric frustration of antiferromagnetic interactions in a kagome network.\cite{norman2016colloquium}  Similarly, in metals, the same kagome tiling can lead to interference effects\cite{PhysRevB.78.125104,kiesel2012sublattice} whose details depend on the position of the Fermi level relative to singularities in the electronic band structure.
	
	Specifically, kagome networks generate electronic structures that are known to host bands with particle-hole asymmetric saddle points at electron fillings $f=5/12$ and $f=3/12$ on either side of Dirac crossings at $f=1/3$ as well as a flat band feature. \cite{wang2013competing,kiesel2013unconventional,wen2010interaction} A form of kinetic frustration due to hopping interference leads to the formation of the localized, flat band that can promote electron-electron interactions. Similarly, for fillings at the saddle points, Van Hove singularities (VHS) result where long-range Coulomb interactions can be promoted due to sublattice interference effects.\cite{kiesel2012sublattice}   These interference effects can generate extraordinarily rich theoretical phase diagrams containing predictions of bond density wave order,\cite{wang2013competing,kiesel2013unconventional} orbital magnetism,\cite{PhysRevB.62.4880, wen2010interaction, lin2021complex}, pair density wave order,\cite{wu2023sublattice} topological insulator phases,\cite{wen2010interaction,PhysRevB.80.113102} and unconventional superconductivity.\cite{PhysRevB.85.144402, kiesel2013unconventional, PhysRevLett.127.177001, PhysRevB.106.L060507}
	
	Recently, considerable interest emerged in finding materials that manifest many of the unconventional electronic states predicted in kagome metals at variable fillings. This involves the identification and study of reasonably two-dimensional metals whose Fermi surfaces/low energy properties are dominated by the electrons occupying their kagome sublattices.  One focus is to identify flat band features and systematically find chemistries/perturbations that bring them toward the Fermi level. Notable recent successes observing unusual correlation effects have been reported om this front.\cite{kang2020topological, ye2021flat, PhysRevLett.128.096601}   A second focus is to search for kagome metals possessing electron-fillings near the saddle points within their band structures, similar to those sought in triangular\cite{PhysRevLett.101.156402} and honeycomb lattices.\cite{Nandkishore2012} 
	
	The VHS accessed at these saddle points generate a logarithmic divergence in the density of states at the M-points (midpoints of the edges) of the Brillouin zone (BZ), and they occur in two different flavors.\cite{kiesel2012sublattice}  The first is the so-called ``p-type” VHS whose wave functions derive from a single sublattice within the three-sublattice kagome network, and the second is the so-called ``m-type” VHS comprised of wave functions mixed between sublattices.  Once the filling corresponding to a VHS is reached, nesting between the three inequivalent M-points across the Fermi surface is predicted to promote a number of charge/spin density wave and superconducting instabilities, with the leading instability dependent on the relative importance of on-site and nearest neighbor Coulomb interactions.\cite{kiesel2013unconventional}
	
	Materials with suitable quasi-two dimensional kagome band structures and fillings near their VHS capable of testing these predictions remained largely elusive until the discovery of the \textit{A}V$_3$Sb$_5$ (\textit{A}=K, Rb, Cs) class of kagome metals.\cite{ortiz2019new}  These compounds were first reported in 2019 as a new structure type built from a vanadium-based kagome network that forms a quasi-two dimensional structure both chemically and electronically.  An important feature is that the kagome lattice of vanadium ions in \textit{A}V$_3$Sb$_5$ is ``nonmagnetic”, meaning that the electrons on the kagome network are delocalized with no local moments (i.e. they are Pauli paramagnets).\cite{kenney2021absence,PhysRevLett.125.247002}  This removes energetically favored local moment magnetism that either outcompetes or masks many of the instabilities predicted in a number of kagome lattice Hubbard models.  A trivial example is the competition between static magnetic order and superconductivity or the dominant response of local moments when searching for weaker, orbital magnetism.  
	
	The goal of this review is to synthesize the experimental progress in understanding the electronic states and phase behaviors identified in the new class of \textit{A}V$_3$Sb$_5$ kagome charge density wave superconductors, where states ranging from orbital antiferromagnetism\cite{PhysRevB.104.165136,PhysRevLett.127.217601,lin2021complex} to electronic nematicity\cite{PhysRevB.107.155131} to pair density wave order\cite{chen2021roton} have been proposed.  The key elements of the crystallographic and electronic structures of this materials class are presented first, followed by an overview of the characteristics of the high-temperature charge density wave (CDW) and the low-temperature superconducting (SC) states.  The presence of intermediate energy scales or crossovers in select \textit{A}V$_3$Sb$_5$ variants will be reviewed as well as the current state of experiments perturbing both CDW and SC order parameters via chemical substitution and pressure.  We view the current review as timely due to the coalescence of much of the experimentally delineated phenomenology surrounding the CDW and SC order parameters in \textit{A}V$_3$Sb$_5$ compounds, and we hope to help focus future measurements probing the microscopic origins of their unconventional electronic properties.
	
	
	\begin{figure}[ht]
		\centering
		\includegraphics[width=\linewidth]{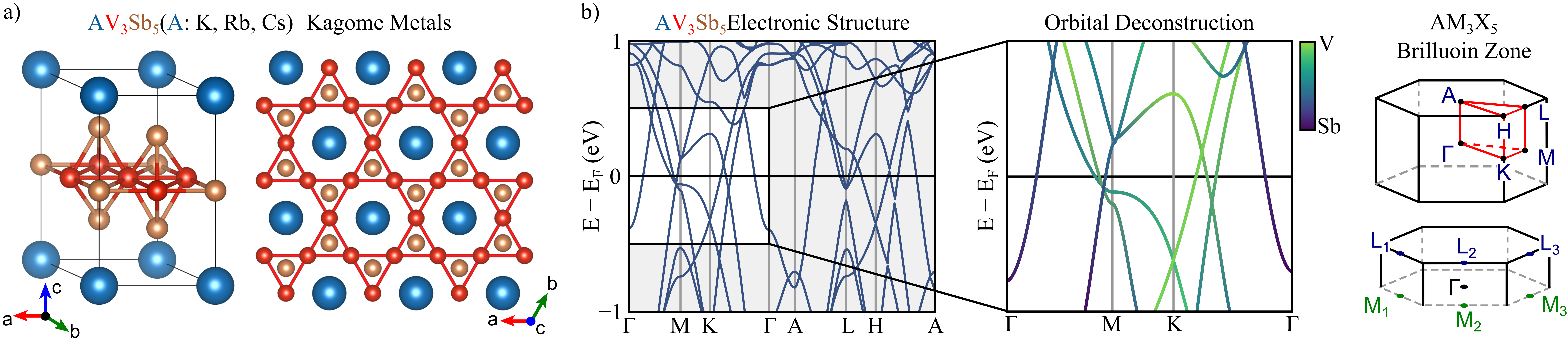}
		\caption{\textbf{Crystal and electronic band structures of $A$V$_3$Sb$_5$ compounds} a | Lattice structure of $A$V$_3$Sb$_5$ with $A$=K, Rb, Cs.  Red spheres show the kagome net of V atoms, each coordinated by an octahedra of Sb atoms depicted as gold spheres.  Between the V$_3$Sb$_5$ layers is a honeycomb lattice of alkali metal $A$-site atoms, depicted as blue spheres.  b | The electronic band structure of $A$V$_3$Sb$_5$ determined via density functional theory calculations.  Key features native to the kagome network are highlighted in the $k_z=0$ plane such as a series of two saddle points with V orbital character just below $E_F$ at the M-point, V-based Dirac points below $E_F$ at the K-point, and a mixed (V,Sb) character saddle point above $E_F$ at the M-point. A representative Brillouin zone (BZ) is also illustrated with the location of high-symmetry points labeled.}
		\label{fig:Structure}
	\end{figure}
	
	\section*{Lattice and electronic structures of $A$V$_3$Sb$_5$}
	
	
	\subsection*{Lattice structure}

	Among the currently popular kagome metal families (e.g. CoSn\cite{CoSn_yu2011near,CoSn_xie2021spin,CoSn_bak1980theory,CoSn_kang2020dirac,CoSn_meier2020flat,CoSn_liu2020orbital,CoSn_sales2021tuning,CoSn_teng2023magnetism,CoSn_kang2020topological,CoSn_sales2019electronic}, FeMn$_6$Ge$_6$\cite{166_arachchige2022charge,166_el1991magnetic,166_ghimire2020competing,166_lee2022anisotropic,166_peng2021realizing,166_pokharel2021electronic,166_pokharel2022highly,166_wang2021field,166_yin2020quantum,166_zhang2022electronic}), the \textit{A}V$_3$Sb$_5$ prototype structure is somewhat distinct, as the V-Sb covalent network that forms the hallmark kagome network is intercalated by a honeycomb network of alkali metal ions. Figure \ref{fig:Structure} shows the normal state crystal structure of the \textit{A}V$_3$Sb$_5$ family. The orthographic perspective shows all bonds with $d<3$ \AA, highlighting the covalent V-Sb sheets and alkali intercalant. The top-down perspective shows only the V-V bonds to highlight the kagome lattice. There are two distinct sublattices of antimony in the \textit{A}V$_3$Sb$_5$ system, and the Sb atom sitting within the kagome network is symmetrically distinct from the Sb atoms above/below the kagome plane. 
	
	The well-isolated kagome sheets are one of the most intriguing features of the \textit{A}V$_3$Sb$_5$ family, and the resulting quasi-two dimensional nature most obviously manifests in the mechanical properties. Single crystals are highly exfoliable, and experiments exploring thickness-dependent measurements have been pursued in parallel to studies of bulk crystals,\cite{thinsong2021competing,thinwei2022linear,thinwu2022nonreciprocal} with recent reports claiming to have stable flakes of CsV$_3$Sb$_5$ thinned to 5 monolayers.\cite{thinsong2021competition} Such facile exfoliation has made the \textit{A}V$_3$Sb$_5$ family a popular target for surface sensitive probes such as scanning tunneling microscopy (STM) and angle-resolved photoemission (ARPES).
	
	The alkali-atom ``intercalation'' in the \textit{A}V$_3$Sb$_5$ is not wholly without comparison, as other alkali metal prototypes (e.g. CsCu$_3$S$_2$,\cite{klepp1996crystal,burschka1980cscu} K$_3$Cu$_3$P$_2$,\cite{savelsberg1978darstellung} Cs$_2$Pd$_3$S$_4$\cite{bronger1971cs2pd3s4}) also exhibit kagome layers separated by alkali atoms.\cite{} The recently reported families of \textit{Ln}V$_3$Sb$_4$ and \textit{Ln}Ti$_3$Bi$_4$ kagome metals also feature slabs of V/Ti kagome networks ``intercalated'' by rare-earth zig-zag chains.\cite{PhysRevMaterials.7.064201,ortiz2023evolution,ovchinnikov2018synthesis,ovchinnikov2019bismuth} However, many of these layered cousins suffer from more complex unit cells, orthorhombic distortions, or increased air sensitivity. Remarkably, despite the layered structure and alkali metal intercalant, the \textit{A}V$_3$Sb$_5$ family is highly tolerant of air, water, and common solvents, increasing the overall accessibility of experiments using single crystals.
	
	Together with the quasi-2D mechanical and structural properties, the reduced dimensionality is critical for the realization of the prototypical ``kagome'' electronic structure.
	Many other candidate kagome metals are derivatives of the CoSn family (e.g. CoSn, GdCo$_3$B$_2$, FeMn$_6$Ge$_6$) or Laves (e.g. MgZn$_2$) prototypes, and maintain a considerable degree of three-dimensional bonding between adjacent kagome layers. Recent computational surveys have highlighted the importance of local bonding and dimensional isolation on the potential to realize the hallmark features of a kagome metal, including the saddle points, Dirac points, and flat bands.\cite{jovanovic2022simple} 
	
	\subsection*{Electronic structure}
	A schematic showing the representative band structure of $A$V$_3$Sb$_5$ compounds is shown in Figure 1 (b).  Across the series of compounds, the band structures are qualitatively similar to one another with bands at the Fermi level dominated by states arising from the kagome nets of vanadium $d$-states. These are multiorbital materials with $d_{xy}$ $d_{yz}$ and $d_{xz}$ derived bands forming a series of m-type and p-type VHS at the M-points of the BZ and at energies reasonably close to $E_F$.\cite{PhysRevLett.127.046401,kang2022twofold, hu2022rich}  Similarly, a Dirac crossing at the K-point of the BZ appears close to $E_F$ in all three compounds.  
	
	While electron-phonon coupling likely also plays a role,\cite{Luo2022, kaboudvand2022fermi, PhysRevB.105.L140501, PhysRevB.106.L081107} well-defined nesting at the M-points in the $k_z=\pi$ plane was identified in ARPES measurements.\cite{kang2022twofold} In CsV$_3$Sb$_5$, states identified with the m-type $d_{xz,yz}$ VHS are nearly perfectly nested and gapped below the CDW transition.  Optical conductivity data resolve the partial gap that opens below the CDW transition to be $\Delta_{CDW}\approx60$ meV in KV$_3$Sb$_5$ with $T_{CDW}=78$ K ,\cite{ece2022optical} $\Delta_{CDW}\approx78$ meV in CsV$_3$Sb$_5$ with $T_{CDW}=94$ K,\cite{PhysRevB.104.045130} and $\Delta_{CDW}\approx100$ meV in RbV$_3$Sb$_5$ with $T_{CDW}=104$ K,\cite{PhysRevB.105.245123}.  This is notably larger than the CDW gaps estimated in surface sensitive STM and ARPES measurements,\cite{kang2022twofold, PhysRevX.11.031026} potentially due to matrix element and surface termination effects.\cite{Huai_2022} 
	
	The relative importance of the multiple saddle points close to $E_F$ remains an important area of study; in particular, whether one or multiple VHS are required to capture the essential physics of the CDW state.  The ordering of the saddle points below $E_F$ switches due to the modified (expanded) lattice of CsV$_3$Sb$_5$ relative to its K- and Rb-based cousins. In DFT models, this relative ordering VHS depends on the interlayer spacing, with KV$_3$Sb$_5$ and RbV$_3$Sb$_5$ showing a configuration with $d_{xz}$/$d_{yz}$-character VHS closest to E$_F$ in the $k_z$=0 plane while CsV$_3$Sb$_5$ inverts the order with $d_{x^2-y^2}$/$d_{z^2}$/$d_{xy}$-character VHS closest to $E_F$.\cite{PhysRevB.104.045130,PhysRevB.104.205129}  This seemingly coincides with a unique CDW ground state in CsV$_3$Sb$_5$\cite{PhysRevMaterials.7.024806} (discussed further in the next section) and suggests that a switch in the orbital character of the Van Hove points closest to $E_F$ impacts the favored charge instability.    
	
	While the VHS arising from the kagome sublattice are important for the stabilization of CDW order, additional states at $E_F$ originating from Sb $p$-orbitals likely also play a role.  The large electron pocket at the $\Gamma$-point is generated by $p$-orbitals from the Sb sites in the kagome plane (in the centers of the hexagons of the kagome nets), and an M-point VHS of mixed Sb/V character appears slightly above $E_F$ derived from a mixture of V-states with out-of-plane Sb $p$-states.\cite{PhysRevB.105.235145}  Doping and pressure-based studies described later in this review have shown that removal of the $\Gamma$-centered Sb band coincides with the suppression of superconductivity and that small, orbital-selective doping of the Sb bands can dramatically renormalize the CDW state.  This phenomenology illustrates that both the VHS from the V $d$-states in the kagome sublattice as well as Sb $p$-states are necessary for minimal models of the physics of these compounds, where, for instance, the Sb $p$-states are proposed to mediate the three-dimensional stacking of the CDW order.\cite{PhysRevB.107.205131,PhysRevB.108.075102}
	
	Quantum oscillation measurements confirm the quasi-two dimensional nature of the vanadium bands,\cite{PhysRevX.11.041030, PhysRevLett.127.207002, PhysRevB.105.024508, PhysRevB.107.155128, PhysRevB.107.075120} and, in the distorted state, a number of low-frequency, CDW-induced vanadium orbits are known to carry a non-trivial Berry phase.\cite{PhysRevB.105.024508, PhysRevLett.127.207002} These orbits have small effective masses consistent with their originating from partially gapped Dirac bands centered at the K-points, and the role of these nontrivial bands in the superconducting phase remains an open question.  Due to the rapid damping of quantum oscillations with increased temperature, measurements of quantum oscillations are largely confined to deep within the CDW state, and more detailed insights are often complicated by the large number of extremal orbits that appear in the reconstructed CDW state.\cite{PhysRevLett.129.157001}
	
	A final, salient point regarding the band structure of \textit{A}V$_3$Sb$_5$ is that there exists a continuous, direct band gap and a series of topological bands at the Fermi level.  This allows for the assignment of a nontrivial ${\mathbb{Z}}_{2}$ invariant and the classification of these compounds as ${\mathbb{Z}}_{2}$ metals hosting topologically nontrivial surface states.\cite{PhysRevLett.125.247002}  These surface states are predicted to be very close to $E_F$ \cite{PhysRevX.11.041030} at the M-points, and there are experimental hints that they are pulled down to the Fermi level once the band structure is modified through the CDW transition.\cite{hu2022topological}  While trivial, bulk states at $E_F$ would mask the impact of these surface states on the low energy properties in the normal state, the protected surface states are potentially important within the superconducting phase where the bulk states become gapped and Fu-Kane (connate) models of topological superconductivity may apply.\cite{PhysRevLett.100.096407}   
	
	\section*{Charge density wave order}
	
	\subsection*{Real component of bond centered order}
	
	\textit{A}V$_3$Sb$_5$ compounds all exhibit CDW order below $T_{CDW}$ = 78, 104, and 94 K for \textit{A} = K, Rb, and Cs respectively. Though it is difficult to resolve directly, the CDW state is predominantly modeled as deriving from bond-centered order. The main observable is a weak structural distortion of the vanadium sublattice \cite{PhysRevLett.125.247002, PhysRevX.11.041030, PhysRevX.11.031050} that maps into energetically favored breathing modes of the kagome plane.\cite{PhysRevLett.127.046401}  This distortion is accompanied by a modulation in the local density of states as imaged via STM measurements.\cite{Jiang2021,Zhao2021,PhysRevB.104.035131}  A number of initial experimental reports identified that the charge density wave state was three-dimensional with a well-defined phasing between neighboring kagome planes,\cite{PhysRevX.11.031026,PhysRevX.11.031050,PhysRevX.11.041030} and the in-plane distortion can be characterized by a 3\textbf{q} breathing mode into a "Star-of-David" (SoD) or "Tri-Hexagonal" (TrH) pattern.  
	
	Both SoD and TrH patterns are supported by \textit{ab intio} calculations modeling the energetically favored distortions of the kagome plane, with the favored distortions being comprised of M-point modes combined with L-point modes mixing in an out-of-plane modulation to the CDW pattern.\cite{PhysRevLett.127.046401,PhysRevB.104.214513,PhysRevMaterials.6.015001}  The result is a staggering of SoD and TrH distortions along the $c$-axis by shifting one pattern of distortion by half an in-plane lattice constant relative to its neighbors.  This staggering along $c$ breaks the in-plane rotational symmetry and results in an orthorhombic unit cell.\cite{PhysRevMaterials.7.024806}  While the relative energies of different distortion types are very close, the commonly predicted distortion mode is the staggered TrH arrangement, comprised of 3\textbf{q}=(M, L, L) modes.
	
	Experimentally, the patterns of charge density wave order differ across the $A$V$_3$Sb$_5$ parent compounds. 
	KV$_3$Sb$_5$ and RbV$_3$Sb$_5$ share a common staggered TrH distortion,\cite{PhysRevMaterials.7.024806,PhysRevResearch.5.L012017,Kang2023} while CsV$_3$Sb$_5$ seemingly possesses a more complex mixture of TrH and SoD layers staggered relative to one another.\cite{PhysRevB.106.L241106,PhysRevB.105.195136,PhysRevMaterials.7.024806,PhysRevResearch.5.L012032}  The average V-V distance ($\approx 2.7$~\AA) is the same in the low-temperature charge density wave state of all three compounds, while the room temperature V-V distance expands with the alkali metal cation size.  This effect combined with the different ordering of the VHS types near $E_F$ seemingly drives a distinct pattern of CDW order in CsV$_3$Sb$_5$ marked by metastability.  Specifically, the out-of-plane modulation of TrH/SoD stacking varies as a function of disorder and thermal history, with regions of $2\times2\times4$ supercells competing with smaller $2\times2\times2$ supercell regions.\cite{PhysRevResearch.5.L012032}  The onset of each of these regions is staged as a function of cooling into the CDW; however, in scenarios where only $2\times2\times2$ order is isolated, then a staggered TrH CDW state can be determined (similar to the K- and Rb- variants).\cite{PhysRevB.105.195136,PhysRevMaterials.7.024806} The experimental observation of mixed TrH and SoD order in the average structure then presumably arises from the $2\times2\times4$ regions of mixed-state crystals.
	
	\begin{figure}[t]
		\centering
		\includegraphics[width=\linewidth]{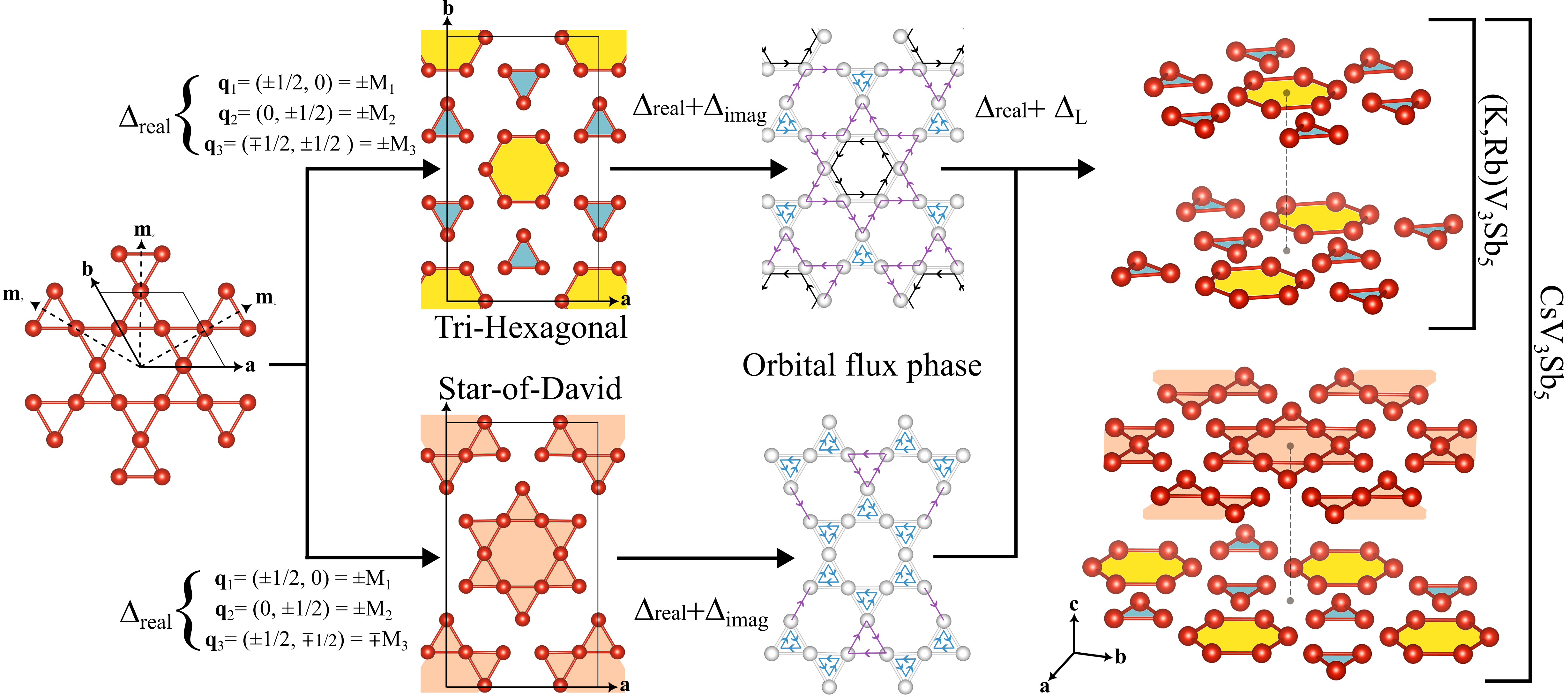}
		\caption{\textbf{Elements of CDW order in \textit{A}V$_3$Sb$_5$ compounds.} The structure of CDW order in $A$V$_3$Sb$_5$ can be thought of a combination of a several key elements.  The first element is the primary component of the real order parameter and is the in-plane 3\textbf{q} distortion of the kagome plane, which is favored as breathing into SoD or TrH-type distortions.  The second element is the proposed imaginary component, which modulates hopping into an orbital flux phase and breaks TRS.  The third element is the interplane correlation that modulates the real component of the CDW state along the $c$-axis.  This arises via consideration of out-of-plane momenta (along the L-points) that modulate the phasing or distortion types between the planes.  RbV$_3$Sb$_5$ and KV$_3$Sb$_5$ each show a TrH in-plane distortion that is staggered by half an in-plane lattice constant along the $c$-axis.  CsV$_3$Sb$_5$ has a mixed-phase CDW, whose average 4-layer structure refines to a mixture of staggered TrH distortion interwoven with staggered TrH and SoD distorted layers.}
		\label{fig:CDW}
	\end{figure}
	
	\subsection*{Imaginary component of bond centered order }
	For band fillings close to a p-type VHS, an imaginary CDW state is predicted to stabilize.  This is effectively a bond-centered CDW that modulates hopping across the kagome network, creating a form of orbital antiferromagnetism that breaks time-reversal symmetry (TRS).  This purely orbital magnetic state exists in the absence of local spins\cite{PhysRevB.104.035142} though it breaks the same symmetries as a spin-density wave. In \textit{A}V$_3$Sb$_5$ compounds, initial hints of TRS-breaking within the CDW state were reported in STM measurements.\cite{Jiang2021,PhysRevB.104.035131,PhysRevB.104.075148}  Here, a Fourier transform of the local density of states showed varying weights at the three inequivalent M-point charge superlattice peaks.  The application of a magnetic field is reported to switch the relative ordering of the Fourier weights of these peaks (or the effective winding between points), suggesting a chiral CDW state that breaks TRS---an inference stemming via its coupling to an external magnetic field. While the same behavior was reported in all three \textit{A}V$_3$Sb$_5$ variants, there is ongoing debate regarding the repeatability of the effect and its origin.  Namely, separate STM studies have failed to resolve a similar magnetic field switching,\cite{Li2022_NatPhys,Li2023,PhysRevB.105.045102} and there is a debate regarding the surface conditions necessary to resolve the effect versus systematic errors in the measurement protocol.  A similar debate exists in optical studies. Here Kerr rotation and circular dichroism initially reported the onset of TRS breaking below the CDW transition of all three $A$V$_3$Sb$_5$ variants;\cite{Xu2022} however subsequent polar Kerr measurements using a Sagnac interferometer failed to resolve net ferromagnetism or \textbf{q}=0 order.\cite{PhysRevLett.131.016901} 
	
	\begin{figure}[t]
		\centering
		\includegraphics[width=\linewidth]{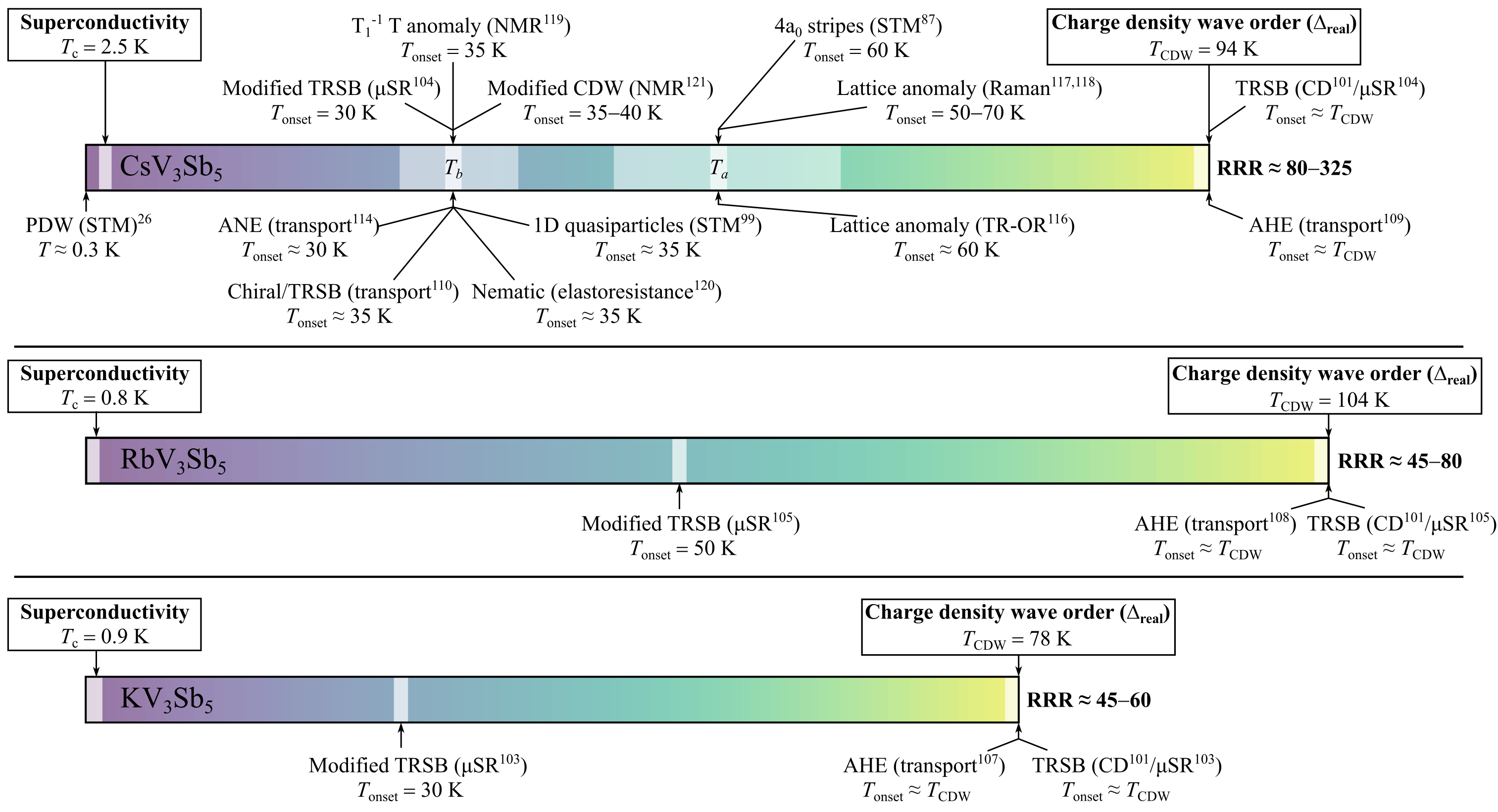}
		\caption{\textbf{Intermediate electronic phase transitions and crossovers in \textit{A}V$_3$Sb$_5$.} The progression of phase transitions and reports of symmetry lowering, such as time reversal symmetry breaking (TRSB), in $A$V$_3$Sb$_5$ compounds.  Reports of lattice and electronic anomalies are visually depicted here and described further in the text.  The range of reported residual resistivity ratios (RRR $\equiv \rho_{4 K}/\rho_{300 K}$) for each compound is summarized next to each chart of anomalies. The majority of anomalies intermediate between the onset of CDW order and SC are currently reported in CsV$_3$Sb$_5$.}
		\label{fig:Instability}
	\end{figure}
	
	In contrast to STM and optical measurements, there is remarkable agreement in muon spin relaxation ($\mu$SR) studies suggesting TRS breaking within the CDW state.\cite{Mielke2022,PhysRevResearch.4.023244,Guguchia2023,yu2021evidence}  This appears in the form of an order-parameter-like modification in the muon spin relaxation rate suggestive of the appearance of a weak, potentially inhomogenous local magnetic field.  The onset of this local field appears below the CDW transition; however, in CsV$_3$Sb$_5$ and RbV$_3$Sb$_5$, there are further modifications observed at lower temperatures.  Notably, the magnitude of the local field is small---comparable to nuclear moments in the sample---and traditional oscillations in the muon spin polarization, indicative of long-range magnetic order, are absent.  This makes interpretation of the microscopic details driving the depolarization challenging; however, the systematics of the effect, its magnetic field dependence, and the anisotropy of the response strongly suggest that they arise from a subtle TRS-breaking effect.
	
	Magnetotransport measurements are another probe suggesting TRS-breaking in the CDW state. An initial measurement of a large, low-field anomalous Hall effect suggested the onset of magnetic order or spin freezing below the CDW transition in KV$_3$Sb$_5$\cite{doi:10.1126/sciadv.abb6003}, and this same effect was subsequently observed in all \textit{A}V$_3$Sb$_5$ variants.\cite{Wang_2023_RVS, PhysRevB.104.L041103}  This is indicative of the presence of a large Berry flux generated upon entering the CDW state; however, crucially, there is no spontaneous ($\mu_0 H=0$ T) component to the anomalous Hall response. The missing zero-field anomalous Hall response is consistent with the absence of a \textbf{q}=0 component of orbital magnetization, and magnetochiral transport studies in CsV$_3$Sb$_5$ further suggest field-switchable chirality, rooted either in the structure or broken TRS.\cite{Guo2022}  Recent torque magnetometery data\cite{asaba2023evidence} also report TRS-breaking albeit with an onset temperature higher than $T_{CDW}$, potentially reflective of the fluctuating/short-range CDW correlations reported at higher temperatures in this compound.\cite{Subires2023,PhysRevLett.129.056401}
	
	\section*{Staged electronic order}
	
	\begin{wrapfigure}{R}{0.5\textwidth}
		\begin{center}
			\includegraphics[width=0.48\textwidth]{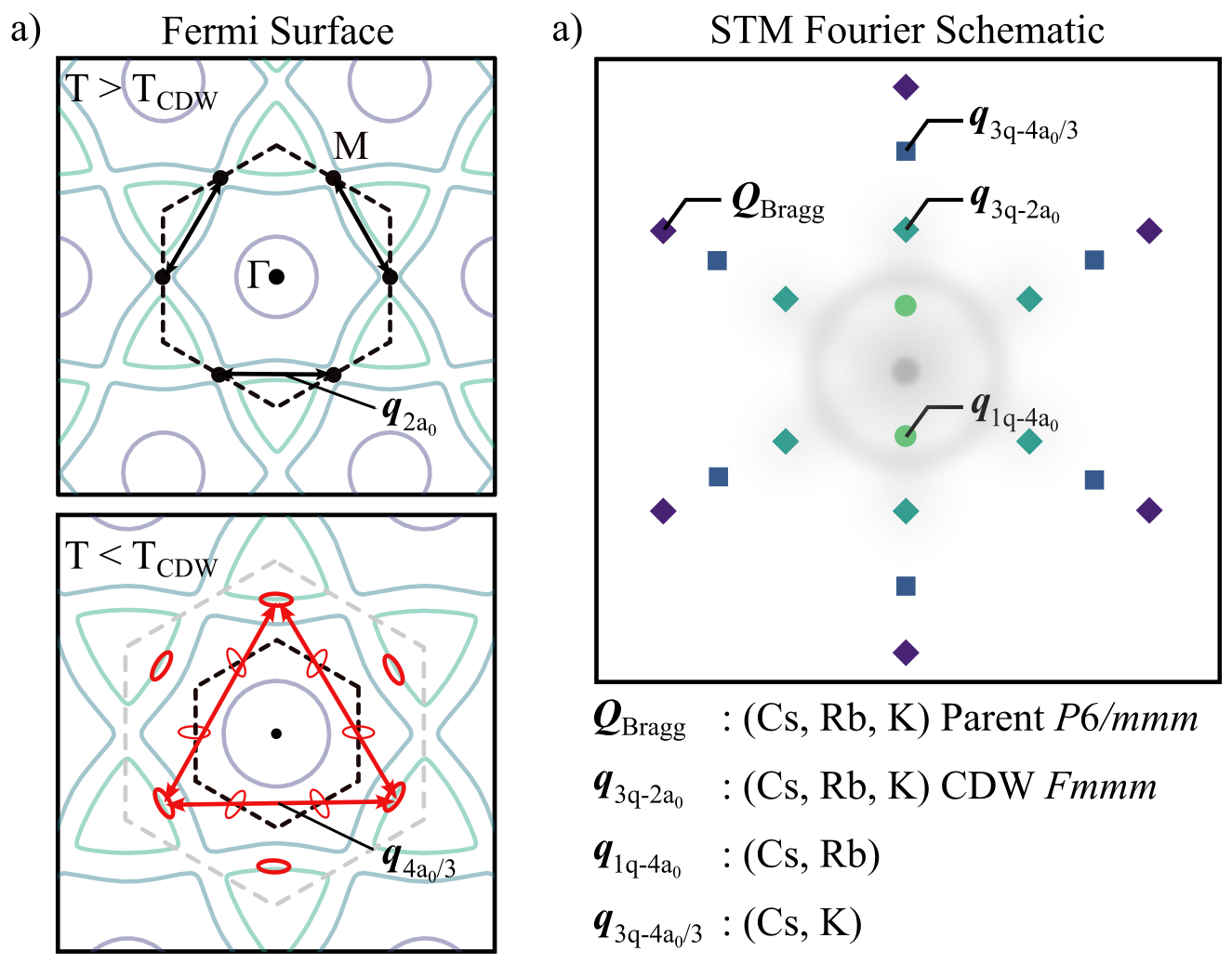}
			\caption{\textbf{Schematic of momentum space contours of the Fermi surface and corresponding nesting wave vectors for electronic order.} a | Nesting wave vectors in the unfolded and folded BZ above and below the CDW transition respectively. Nested M-points are illustrated in the unfolded zone while a schematic of nested Chern pockets are highlighted in the folded zone, below $T_{CDW}$. b | Wave vectors of charge correlations resolved within the $ab$-plane as reported via STM measurements and a a tabulation of compounds where these correlations have been reported.}
		\end{center}
		\label{fig:Fermisurface}
	\end{wrapfigure}
	
	A further signature of the unconventional nature of CDW order in  \textit{A}V$_3$Sb$_5$ compounds is the appearance of a number of intermediate crossover behaviors below $T_{CDW}$ upon cooling toward the SC phase.  Figure 3 illustrates the temperature scales of these anomalies, suggesting a staged evolution of electronic order upon cooling.  The vast majority of these reports are reported in  CsV$_3$Sb$_5$, with only muon spectroscopy reporting intermediate states in the RbV$_3$Sb$_5$ and KV$_3$Sb$_5$ compounds.  
	
	This difference between compounds likely stems from one of two origins.  The first is that the crystal quality of CsV$_3$Sb$_5$ is typically superior to that of the other variants, with residual resistivity ratios (RRR) reported as high as 300 (compared to RRR=60-80 reported in the Rb- and K-variants).\cite{PhysRevB.105.L201109,PhysRevB.105.094507,PhysRevB.107.075120}  This engenders greater exploration by the community, and the lower disorder potentially stabilizes or unmasks subtle experimental signatures of electronic staging in the CDW state. In this scenario, staging of the electronic order is naively present in all three \textit{A}V$_3$Sb$_5$ compounds, but it remains hidden in less frequently studied crystals with poorer quality.   
	
	The second possibility is that the subtly different band structure of CsV$_3$Sb$_5$ and the change in the relative ordering of VHS close to $E_F$ generates this staging effect.  This scenario would imply that the distinct pattern of CDW order and metastability that CsV$_3$Sb$_5$ realizes creates a distinct thermal evolution of the CDW state upon cooling.  As will be discussed later, CsV$_3$Sb$_5$ also possesses distinct doping- and pressure-tuned phase diagrams, suggesting the this second scenario of a unique starting CDW state is the most likely origin for its richer thermal evolution of electronic states.
	
	Focusing on CsV$_3$Sb$_5$, a number of experimental probes report signatures of either lattice or electronic anomalies near $T_a\approx60$ K and $T_b\approx35$ K as summarized in the top panel of Figure 3.  Each of these temperature scales lacks a sharp thermodynamic anomaly in the heat capacity, suggesting they originate from a subtle crossover in the electronic structure. The first anomaly $T_a$ is defined primarily by the emergence of a short-lifetime phonon mode in optics measurements,\cite{PhysRevMaterials.5.L111801,PhysRevB.105.155106,PhysRevResearch.4.023215} suggesting that it is coupling to an electronic degree of freedom.  $T_a$ also coincides with the appearance of quasi-one dimensional charge stripes with a real space lattice modulation of four lattice constants ($4a_0$) on the surface resolved by STM studies.\cite{Zhao2021}  Upon further cooling, a second energy scale appears at $T_b$.  This energy scale is characterized by probes reporting rotational symmetry breaking and higher harmonics in magnetotransport, suggestive of the onset of chirality or time reversal symmetry breaking.\cite{Song2022,Nie2022,Guo2022,PhysRevB.105.L201109}  Notably, $T_b$ also coincides with the onset of quasi-one dimensional band features in quasi-particle interference spectra\cite{Li2023} as well as a modification in the local Sb environment\cite{Luo2022_2} and changes in the local magnetic field resolved in muon spin relaxation studies.\cite{PhysRevResearch.4.023244}  One potential interpretation is that the $T_a$ energy scale represents the onset of emergent CDW fluctuations that couple to the lattice and slowly freeze toward $T_b$, affecting a crossover in the electronic structure and transport properties.
	
	
	One possible driver of staged behavior within the CDW state is the CDW-driven nesting of small pockets in the folded BZ following the onset of CDW order. Figure 4 (a) illustrates one possible scenario where the nested VHS at the M-points in the folded, $P6/mmm$ cell reconstruct the zone via a TrH distortion into a smaller BZ.  The reconstructed zone possesses small pockets that nest along a new \textbf{q}=$\frac{3}{2}$M (and equivalent wave vectors) and can drive a secondary instability at lower temperature.\cite{Zhou2022} These small pockets were recently observed in a joint STM and ARPES study, and they are proposed to be Chern pockets that support nesting for a new 3\textbf{q}-type order with a real space modulation of ($4a_0/3$).\cite{PhysRevX.13.031030}
	
	The proposed new 3\textbf{q} wave vector corresponds to the anomalous charge correlations that appear at low-temperature and modulate the superfluid density at the surface as reported in scanning Josephson tunneling measurements.\cite{chen2021roton}  This observation has invoked the notion of an intertwined CDW and SC state connected through a primary pair density wave instability.\cite{wu2023sublattice}  How ubiquitous this phenomenon is in other \textit{A}V$_3$Sb$_5$ compounds, and whether the emergent $4a_0/3$-type correlations arise from the nearby $T_b$  energy scale remains to be established.  Figure 4 (b) summarizes the reports of local charge correlations present in the different material classes and their visualization in STM measurements.
	\newline
	
	\section*{Superconducting order}
	
	All three parent $A$V$_3$Sb$_5$ compounds host a superconducting transition within the CDW state, with the highest $T_c=2.5$ K in CsV$_3$Sb$_5$\cite{PhysRevLett.125.247002} and $T_c\approx0.9$ K for (Rb,K)V$_3$Sb$_5$.\cite{Yin_2021,PhysRevMaterials.5.034801}  Note that the precise $T_c$ reported for each compound varies somewhat between experimental reports (some higher and some lower by a few hundred mK). Due to the competition between SC and the CDW states under slight doping, vetting the ``correct" $T_c$ requires detailed parametrization of the CDW state in the same sample. The SC state forms in the clean limit with $l>>\xi_{ab}$ (where $l$ is the mean free path and $\xi$ is the coherence length),\cite{Duan2021} and the SC state is highly anisotropic with a critical field ratio $H_{c2,c}/H_{c2,ab}\approx9$.\cite{Ni_2021}  
	
	One of the main challenges in studies of \textit{A}V$_3$Sb$_5$ compounds is to conclusively define the pairing symmetry of the SC order parameter.  Despite a number of initial mixed results reporting nodeless versus nodal SC order parameters, the experimental picture has slowly converged to a nodeless,\cite{Duan2021,Zhong2023} anisotropic\cite{Roppongi2023} SC gap function with singlet pairing\cite{Mu_2021}.  A multiband, two gap SC state manifests where one gap is substantially smaller than the other\cite{PhysRevB.104.174507,Gupta2022}---making conventional assessment of low-temperature thermal transport and ``U"- versus ``V"-shaped SC gap spectra in STM challenging due to quasiparticle contamination from the lower gap.  The lower gap size determined via penetration depth measurements is estimated to be $\Delta_{small}\approx 0.5$ $k_B T_c$ \cite{Roppongi2023,Duan2021} while the larger gap from tunneling measurements is estimated to range between $\Delta_{large}\approx 2.5 - 3.6$ $k_B T_c$.\cite{PhysRevB.104.174507,chen2021roton}
	
	A number of different types of SC are predicted to emerge due to nested VHS in the kagome band structure. A leading instability in many models is for a chiral $d+id$ SC state to emerge.\cite{kiesel2012sublattice}  Such a state would be consistent with the observation of a nodeless SC gap function and reports of broken TRS in the superconducting state;\cite{Guguchia2023} however recent irridation-based studies of the response of the SC gap to disorder suggest that such a state can be precluded (as well as sign changing $s^{\pm}$).\cite{Roppongi2023} This assessment is based on a conventional picture of a sign-changing gap function being more sensitive to lattice disorder; however recent theoretical models of SC on the kagome lattice suggest that this conventional assumption may not be valid.\cite{PhysRevB.108.144508}  Specifically, the sublattice character of the kagome network near Van Hove fillings renders disorder to be non-pairing breaking for singlet pairing mechanisms irregardless of whether there is a sign change in the gap function.  Future work exploring this idea and whether $d+id$ pairing can truly be precluded from existing results is merited.  
	
	Upon warming outside of the saturated SC state, additional anomalies are reported at the high temperature phase boundary in the fluctuation regime. Little-Parks measurements probing magnetoresistance oscillations in SC ring devices report an evolution from 2\textit{e}- to 4\textit{e}- to 6\textit{e}-pairing upon warming through the SC transition (where \textit{e} is the electron charge).\cite{ge2022discovery}  Deep within the SC state, conventional 2\textit{e} oscillations were observed; however these oscillations increase in frequency from 4\textit{e} to 6\textit{e} in a broadened regime of finite resistance. The SC transitions in these devices are substantially broader than those in bulk single crystals, suggesting that disorder/strain imparted during the fabrication process creates an extended fluctuation regime. While these Little-Parks measurements have yet to be replicated, recent mutual inductance measurements report an extended vortex liquid regime in CsV$_3$Sb$_5$ \cite{zhang2023vortex}, suggesting that a similar extended fluctuation regime may be accessible in bulk crystals.  6\textit{e} pairing states are predicted to stabilize in such a fluctuation regime in the presence of orbital antiferromagnetism.\cite{varma2023extended}
	
	\section*{Interplay between superconductivity and charge density wave order}
	\begin{wrapfigure}{R}{0.5\textwidth}
		\begin{center}
			\includegraphics[width=0.5\textwidth]{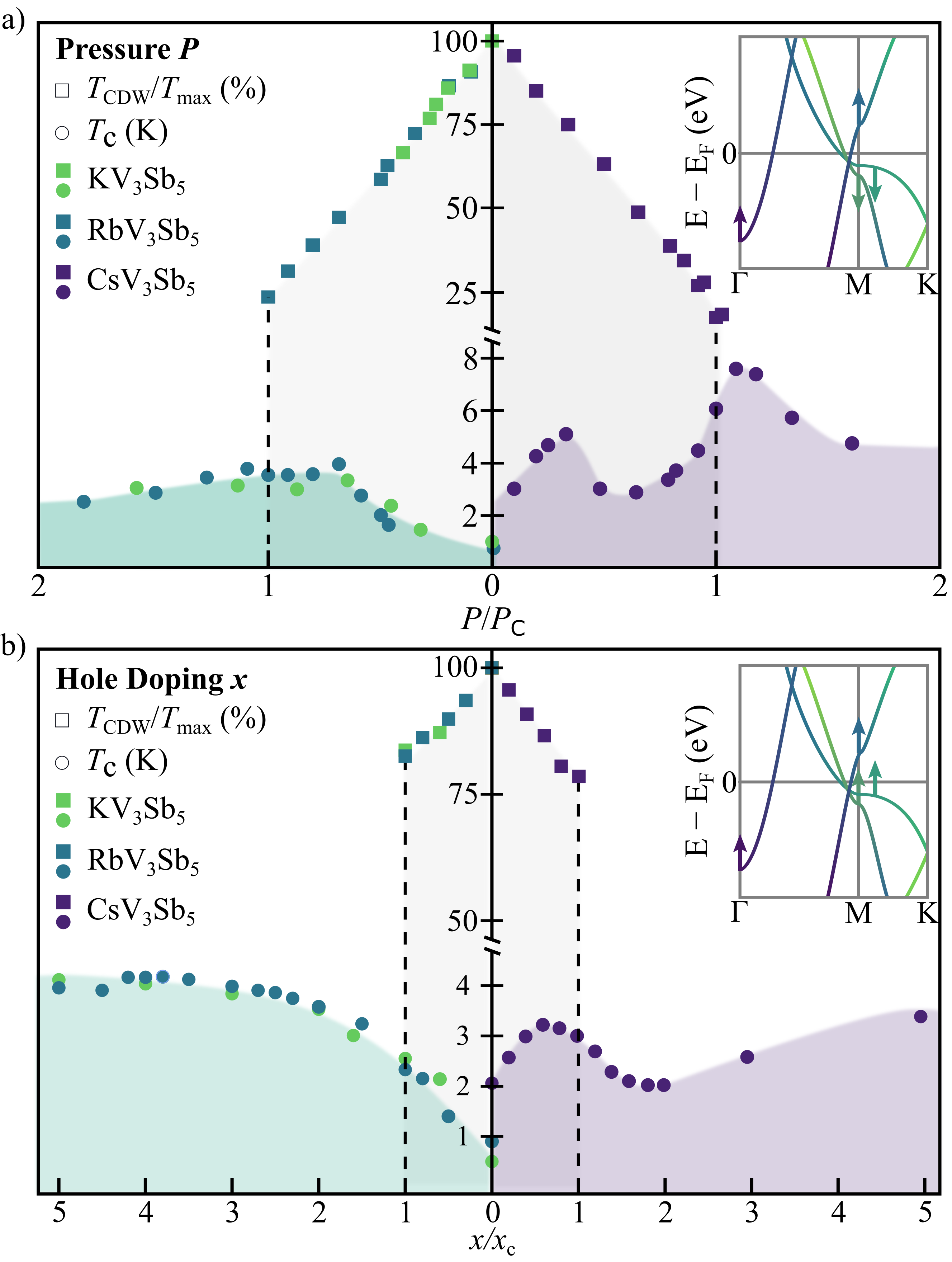}
			\caption{\textbf{Doping and pressure-tuned phase diagrams of $A$V$_3$Sb$_5$ compounds.} a | Pressure-Temperature electronic phase diagram showing the evolution of CDW and SC orders as a function of normalized pressure.  The pressure value is normalized by the critical pressure $P_c$ where the CDW state is reported to vanish.  b | Electronic phase diagram as a function of normalized hole-doping.  Doping concentrations have been normalized by the critical hole-doping value $x_c$ where CDW order nominally vanishes.  CDW transition temperatures have been normalized relative to their undoped, ambient pressure values of 100$\%$. Data were adapted from \cite{PhysRevLett.126.247001,PhysRevResearch.3.043018,PhysRevB.103.L220504,PhysRevMaterials.6.L041801,PhysRevMaterials.6.074802}.}
		\end{center}
		\label{fig:PhaseDiagram}
	\end{wrapfigure}
	The local interplay between charge correlations and the SC state is most extensively visualized in STM measurements, where, at least at the surface and excluding the emergent \textbf{q}=$\frac{3}{2}$-type wave vector, there is smooth coexistence of $2\times2$-type charge correlations and the SC gap. Perturbing the CDW state via external pressure or chemical doping, however, can have a dramatic effect on the SC phase, which we summarize below.
	
	As alluded to earlier in this paper, the pressure and doping responses of KV$_3$Sb$_5$ and RbV$_3$Sb$_5$ differ from those of CsV$_3$Sb$_5$.  For (K, Rb)V$_3$Sb$_5$, the application of hydrostatic pressure rapidly suppresses $T_{CDW}$ while simultaneously enhancing $T_c$.\cite{PhysRevResearch.3.043018,PhysRevB.103.L220504}  There is a continuous trade-off between the two states, where $T_c$ is enhanced as the CDW state is suppressed and frees up a greater density of states for the SC condensate.  The left hand side of Figure 5 (a) shows this common response as the normalized CDW transition temperature for both compounds follows a similar pressure dependence normalized for the critical pressure ($P_c$) necessary to destabilize CDW order.  The suppression of the CDW state eventually terminates in a first-order line where $T_c$ is maximized.  The evolution of the SC state up to and across $P_c$ is nearly identical between the two compounds, and a maximal $T_c=4$ K is realized for both materials once CDW order is suppressed. The conventional trade-off between $T_{CDW}$ and $T_c$ proceeds as pressure naively pushes the Fermi level away from the two occupied VHS closest to $E_F$ as shown in the inset of Figure 5 (a).\cite{PhysRevB.104.205129} 
	
	On the right hand side of Figure 5 (a), a distinct pressure-induced response of CsV$_3$Sb$_5$ is shown using the same normalized critical pressure and CDW onset temperatures.\cite{PhysRevLett.126.247001,Yu2021}  Pressure again drives a rapid suppression of CDW order that terminates in a first-order line; however $T_c$ evolves in a nonmonotonic manner, forming two SC ``domes" within the ($P$, $T$) phase diagram. The first dome reaches a peak $T_c$ \textit{within} the long-range ordered CDW region of the phase diagram.  With continued increase in pressure, the CDW is monotonically suppressed, and $T_c$ decreases to close to its zero-pressure value before increasing again to a global maximum near the first-order CDW phase boundary.  Moving beyond this boundary with increasing pressure causes $T_c$ to decrease again and form a second, extended ``dome".  The termination of the second, higher pressure SC dome seemingly correlates with the removal from the Fermi surface of the Sb $p_z$ states forming the electron-like pocket centered at the $\Gamma$-point of the BZ.\cite{PhysRevB.107.174107}  The lower pressure dome, in contrast, is likely driven by a CDW transition where charge correlations are weakened as they evolve out of the distinct, parent CDW order in CsV$_3$Sb$_5$ into an incommensurate charge density wave state,\cite{Feng2023,Zheng2022} and partial volume fraction SC is reported near this phase boundary.
	
	Tuning the electron-filling (and thus the Fermi level alignment with the VHS) is another means of studying the interplay between the CDW order and its coupling to the SC state. Figure 5 (b) shows the normalized CDW transitions for all three parent compounds as a function of critical concentrations of hole-doping $x_c$ where $x$ is the number of doped holes per formula unit.  As shown in the inset, hole-doping is naively expected to shift the Fermi level closer to occupied VHS in the band structure, though in an orbitally selective manner with the major changes expected in the filling of the Sb-derived $\Gamma$-pocket.\cite{PhysRevB.104.205129,PhysRevMaterials.6.L041801} Again, the relative responses of (K, Rb)V$_3$Sb$_5$ and CsV$_3$Sb$_5$ are distinct. 
	
	Looking first at the left side of Figure 5 (b), hole-doping drives a rapid suppression of CDW order in (K, Rb)V$_3$Sb$_5$, resulting in a common enhancement in $T_c$ up to 4 K.\cite{PhysRevMaterials.6.074802}  On the right hand side of the plot, CsV$_3$Sb$_5$ again shows a double-dome-type evolution where $T_c$ first reaches a maximum \textit{inside} the CDW state followed by a minimum just beyond the first-order phase boundary where CDW order vanishes.\cite{PhysRevMaterials.6.L041801}  Continued hole-doping drives an increase in $T_c$ to a second maximum outside of the CDW state near $x\approx0.33$. While not shown in Figure 5 (b), continued hole-doping then drives a slight decrease in $T_c$ before SC vanishes in the second dome near $x\approx0.7$.  Again, SC vanishes near the doping concentration where the Sb-derived $\Gamma$-pocket is predicted to be lifted above $E_F$ in DFT calculations.  These phase diagrams can be most extensively mapped using Sn-atoms as hole-dopants replacing Sb, and qualitatively similar phase diagrams form using Ti-atoms as dopants replacing V,\cite{Sur2023,yang2022titanium} albeit with lower solubility limits and stronger disorder effects.
	
	There are clear commonalities in the ($P$,$T$) and ($x$, $T$) phase diagrams.  For instance, the rapid suppression of the CDW state for all compounds as a function of pressure and hole-doping is uniformly observed and anomalous, in particular given the differing effects on proximities of the VHS to $E_F$ for the two different types of perturbations.  The differing response of SC to the suppression of CDW order between CsV$_3$Sb$_5$ and (K,Rb)V$_3$Sb$_5$ using both types of perturbations likely arises from the unique starting CDW phase of CsV$_3$Sb$_5$. The charge correlations in the seemingly metastable starting CDW state of CsV$_3$Sb$_5$ change in character under small perturbations, driving strong charge fluctuations in the SC state as they evolve.  The phase boundary or crossover out of the metastable starting CDW state is one possible origin for the initial low-pressure/low-doping dome in CsV$_3$Sb$_5$.  
	
	Incommensurate quasi-one dimensional charge correlations were recently observed near the CDW phase boundary of hole-doped CsV$_3$Sb$_5$ suggesting such a crossover may exist.\cite{Kautzsch2023}  NMR measurements similarly report the presence of incommensurate charge correlations near the boundary between pressure-driven SC domes in this material.\cite{Zheng2022}  Whether or not similar incommensurate charge correlations emerge beyond the CDW phase boundaries of (K,Rb)V$_3$Sb$_5$, however, has yet to be explored.  A second commonality in the pressure/doping phase diagrams of CsV$_3$Sb$_5$ is the complete suppression of SC once the Sb $p_z$ states are driven away from $E_F$, suggesting that these states remain essential to stabilizing SC.  Furthermore, given that the primary effect on the band structure of both hydrostatic pressure and hole-doping is an orbitally selective modification of the Sb $\Gamma$-pocket, the rapid suppression of CDW order in both phase diagrams suggests that the Sb states are intertwined with the V-atom driven CDW order in an unconventional manner.  Resonant x-ray scattering measurements have further resolved that Sb states are coupled/hybridized within the CDW transition\cite{Li2022} despite minimal motion of Sb sites through $T_{CDW}$.

	\section*{Outlook/Future Perspectives}
	
	\begin{wrapfigure}{L}{0.5\textwidth}
		\begin{center}
			\includegraphics[width=0.5\textwidth]{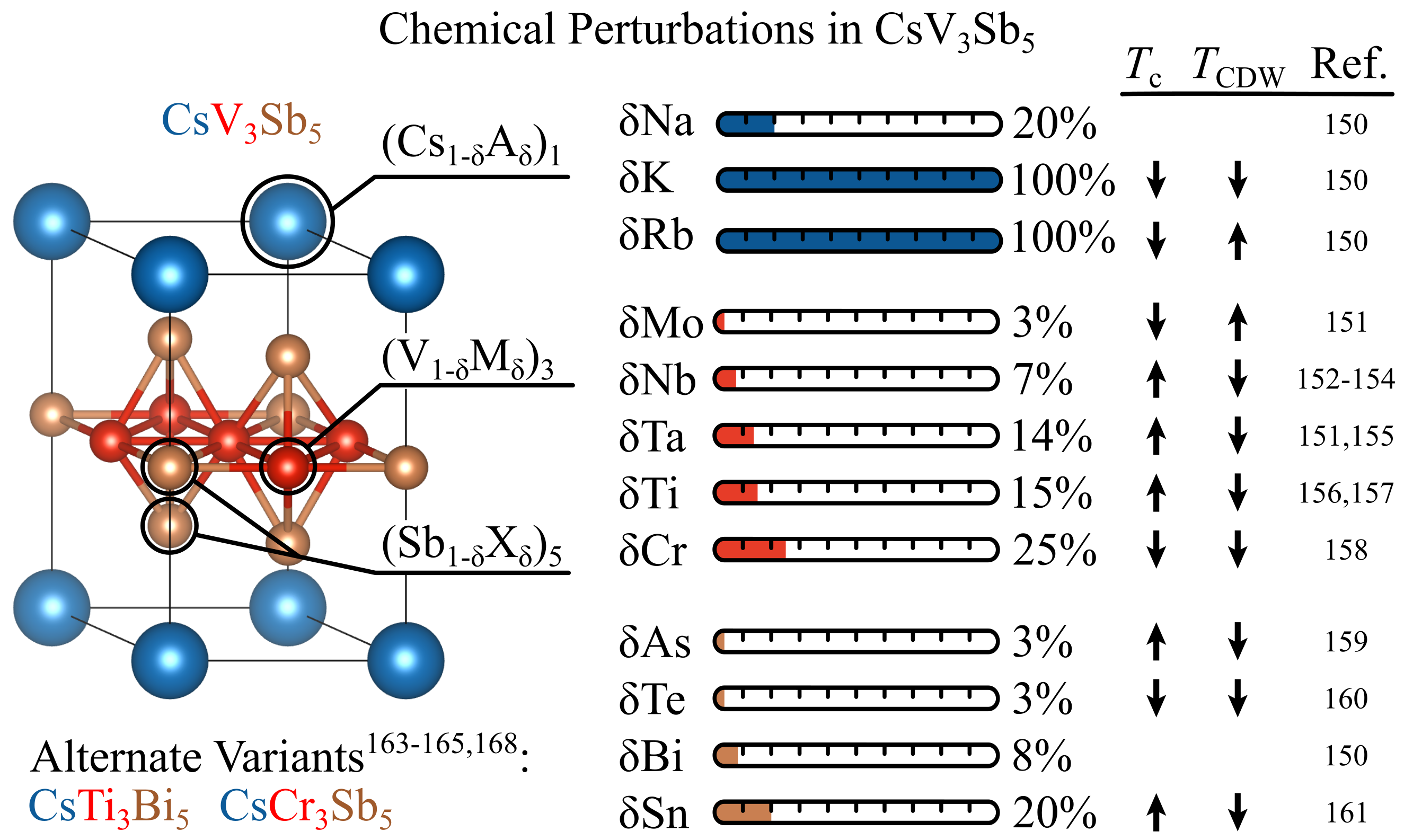}
			\caption{\textbf{Schematic showing sublattice doping of CsV$_3$Sb$_5$ with various electronic, magnetic, and isoelectronic dopants.}  Percent substitution achievable for dopants on each site of CsV$_3$Sb$_5$ and their influence on $T_{CDW}$ and $T_c$ are illustrated. }
		\end{center}
	\end{wrapfigure}
	
	Considerable effort exploring the chemical flexibility of $A$V$_3$Sb$_5$ compounds both in terms of filling control and in terms of isoelectronic/steric perturbation has been reported.\cite{ortiz2023complete,liu2022evolution,li2022tuning,zhou2023effects,xiao2023evolution,li2022strong,liu2023doping,hou2023effect,ding2022effect,liu2022enhancement,capa2023electron,oey2022fermi,yang2022titanium,lei2023band}  A sampling of various chemical substitutions that have been achieved and the corresponding responses of the CDW and SC states for CsV$_3$Sb$_5$ is summarized in Figure 6. Where both polycrystalline and single crystal solubility limits are available, deference is given to single crystal results. Generally speaking, solubility limits seem largest in CsV$_3$Sb$_5$, and by far the most research continues to be invested into the Cs-variant of the structure. Site substitution is possible to varying degrees on all sites in the lattice, which provides a valuable litmus for testing the essential band features and interactions necessary for stabilizing the various types of electronic order in these compounds. 
	
	Crucially, new parent systems with the same structure type have also been discovered, providing a pristine setting for exploring the impact of forming the same kagome lattice at different fillings.  For instance, the recently reported Ti-based variants (Rb,Cs)Ti$_3$Bi$_5$ possess a dramatically different band structure and no signatures of CDW order.\cite{werhahn2022kagome,PhysRevLett.131.026701,Wang_2023}  There are however reports of an intrinsic rotational symmetry breaking in the quasiparticle spectra of these compounds, suggesting a native nematic electronic instability and continued correlation effects.\cite{li2023electronic}  At lower temperatures, superconductivity was reported in CsTi$_3$Bi$_5$, though there exists a debate whether the SC state is intrinsic or arises from an impurity phase.\cite{werhahn2022kagome, yang2022titaniumbased}  A new Cr-based variant CsCr$_3$Sb$_5$ was also very recently reported with a complex evolution of charge order and potential coexisting, local moment magnetic order.\cite{liu2023superconductivity}  These are exciting developments and suggest further unconventional states can be realized via engineering added interactions across the kagome network of the \textit{A}\textit{M}$_3$\textit{X}$_5$ structure-type.
	
	As part of the understanding the origin of the anomalous properties at the band fillings in \textit{A}V$_3$Sb$_5$, the role of the seemingly nested VHS at $E_F$ in driving the staged phase behavior needs to be further constrained experimentally, and the band features necessary for a minimal model of their properties need to be determined.  In particular, the relative importance of the in-plane and out-of-plane Sb $p$-states in stabilizing CDW order, SC, or both is an important open question.  Other material comparators with similar band fillings may provide clues to this.  For instance, kagome net \textit{R}V$_6$Sn$_6$ compounds (\textit{R}=rare earth), while more three-dimensional, possess similar VHS near their Fermi levels, yet they lack similar phase electronic phase transitions.\cite{PhysRevB.104.235139,doi:10.7566/JPSJ.90.124704} A crucial difference is likely that the Sn sites are pushed out of the V-based kagome nets, removing a comparable Sn $p$-pocket at the $\Gamma$-point in the BZ.   
	
	Going forward, crucial experiments directly resolving the symmetries broken in the CDW state of AV$_3$Sb$_5$ are either planned or underway.  Whether or not TRS is broken via a bond-centered CDW is a central question and, if confirmed, would represent the first manifestation of orbital antiferromagnetism in the solid state. One likely resolution to experimental discrepancies in resolving a TRS-broken state is the impact of strain on the response of \textit{A}V$_3$Sb$_5$ compounds. While externally applied, in-plane strain has a muted impact on the relative $T_c$ and $T_{CDW}$ values (likely driven via the Poisson ratio),\cite{PhysRevB.104.144506} strain fields either frozen within crystals or imparted during mounting/cooling samples have recently been shown to have a dramatic impact on the electronic responses associated with TRS- and rotational symmetry breaking.  
	
	Specifically, removal/minimization of strain fields within crystals of CsV$_3$Sb$_5$ seemingly governs whether rotational symmetry breaking is observable within in-plane charge transport measurements.\cite{guo2023correlated} Furthermore, an applied magnetic field orthogonal to the kagome planes induces in-plane transport anisotropy, suggestive of a piezomagnetic response and a natural coupling of TRS-breaking order to strain.   Recent STM measurements directly resolve such a piezomagnetic response as well as optically-induced switching of chirality in the CDW state at the surface of RbV$_3$Sb$_5$ crystals.\cite{xing2023optical}   The emerging picture is then one of a native orbital antiferromagnetic state that breaks TRS and is strongly coupled to out-of-plane magnetic fields and in-plane strain fields.\cite{PhysRevB.106.144504}  These fields can imbalance the components of the multi-q CDW order and induce a net ferromagnetic or \textbf{q}=0 signal detectable by a number of probes (such as Kerr rotation measurements).  Future work exploring the notion of strain's impact on the weak magnetic signal detected in muon spin relaxation and optics measurements is an exciting path forward. 
	
	Resolving the above puzzles will provide crucial hints for the pairing symmetry of the lower temperature superconducting state in AV$_3$Sb$_5$ compounds and hopefully motivate the search for new materials platforms that host similar band structures.  The number of new $A$$M_3$$X_5$ variants recently uncovered is a promising new direction for exploring other correlated states possible on a kagome network. We envision many new opportunities emerging as this new materials phase space is fully explored and the rich frontier of states predicted within kagome metals tuned near their Van Hove fillings can be tested in real materials platforms. 
	\newline
	\newline
	\noindent\textbf{Acknowledgements}\\
	S.D.W. gratefully acknowledges support from the UC Santa Barbara NSF Quantum Foundry funded via the Q-AMASE-i program under award DMR-1906325 and the Eddleman Center for Quantum Innovation. B.R.O gratefully acknowledges support from the U.S. Department of Energy (DOE), Office of Science, Basic Energy Sciences (BES), Materials Sciences and Engineering Division. 
\newline
\newline
	\noindent\textbf{Author contributions}\\
	S.D.W. and B.R.O. composed the manuscript and created the figures.
\newline
\newline
	\noindent\textbf{Competing interests}\\
	S.D.W. and B.R.O. declare no competing financial interests in creating this work.
	Notice: This manuscript has been authored by UT-Battelle, LLC under Contract No. DE-AC05-00OR22725 with the U.S. Department of Energy. The United States Government retains and the publisher, by accepting the article for publication, acknowledges that the United States Government retains a non-exclusive, paid-up, irrevocable, world-wide license to publish or reproduce the published form of this manuscript, or allow others to do so, for United States Government purposes. The Department of Energy will provide public access to these results of federally sponsored research in accordance with the DOE Public Access Plan (http://energy.gov/downloads/doe-public-access-plan).

	\bibliography{sample2.bib}

\begin{thebibliography}{100}
\expandafter\ifx\csname url\endcsname\relax
  \def\url#1{\texttt{#1}}\fi
\expandafter\ifx\csname urlprefix\endcsname\relax\def\urlprefix{URL }\fi
\providecommand{\bibinfo}[2]{#2}
\providecommand{\eprint}[2][]{\url{#2}}

\bibitem{syozi1951statistics}
\bibinfo{author}{Sy{\^o}zi, I.}
\newblock \bibinfo{title}{Statistics of kagom{\'e} lattice}.
\newblock \emph{\bibinfo{journal}{Progress of Theoretical Physics}}
  \textbf{\bibinfo{volume}{6}}, \bibinfo{pages}{306--308}
  (\bibinfo{year}{1951}).

\bibitem{norman2016colloquium}
\bibinfo{author}{Norman, M.}
\newblock \bibinfo{title}{Colloquium: Herbertsmithite and the search for the
  quantum spin liquid}.
\newblock \emph{\bibinfo{journal}{Reviews of Modern Physics}}
  \textbf{\bibinfo{volume}{88}}, \bibinfo{pages}{041002}
  (\bibinfo{year}{2016}).

\bibitem{PhysRevB.78.125104}
\bibinfo{author}{Bergman, D.~L.}, \bibinfo{author}{Wu, C.} \&
  \bibinfo{author}{Balents, L.}
\newblock \bibinfo{title}{Band touching from real-space topology in frustrated
  hopping models}.
\newblock \emph{\bibinfo{journal}{Phys. Rev. B}} \textbf{\bibinfo{volume}{78}},
  \bibinfo{pages}{125104} (\bibinfo{year}{2008}).

\bibitem{kiesel2012sublattice}
\bibinfo{author}{Kiesel, M.~L.} \& \bibinfo{author}{Thomale, R.}
\newblock \bibinfo{title}{Sublattice interference in the kagome hubbard model}.
\newblock \emph{\bibinfo{journal}{Physical Review B}}
  \textbf{\bibinfo{volume}{86}}, \bibinfo{pages}{121105}
  (\bibinfo{year}{2012}).

\bibitem{wang2013competing}
\bibinfo{author}{Wang, W.-S.}, \bibinfo{author}{Li, Z.-Z.},
  \bibinfo{author}{Xiang, Y.-Y.} \& \bibinfo{author}{Wang, Q.-H.}
\newblock \bibinfo{title}{Competing electronic orders on kagome lattices at van
  hove filling}.
\newblock \emph{\bibinfo{journal}{Physical Review B}}
  \textbf{\bibinfo{volume}{87}}, \bibinfo{pages}{115135}
  (\bibinfo{year}{2013}).

\bibitem{kiesel2013unconventional}
\bibinfo{author}{Kiesel, M.~L.}, \bibinfo{author}{Platt, C.} \&
  \bibinfo{author}{Thomale, R.}
\newblock \bibinfo{title}{Unconventional fermi surface instabilities in the
  kagome hubbard model}.
\newblock \emph{\bibinfo{journal}{Physical review letters}}
  \textbf{\bibinfo{volume}{110}}, \bibinfo{pages}{126405}
  (\bibinfo{year}{2013}).

\bibitem{wen2010interaction}
\bibinfo{author}{Wen, J.}, \bibinfo{author}{R{\"u}egg, A.},
  \bibinfo{author}{Wang, C.-C.~J.} \& \bibinfo{author}{Fiete, G.~A.}
\newblock \bibinfo{title}{Interaction-driven topological insulators on the
  kagome and the decorated honeycomb lattices}.
\newblock \emph{\bibinfo{journal}{Physical Review B}}
  \textbf{\bibinfo{volume}{82}}, \bibinfo{pages}{075125}
  (\bibinfo{year}{2010}).

\bibitem{PhysRevB.62.4880}
\bibinfo{author}{Nayak, C.}
\newblock \bibinfo{title}{Density-wave states of nonzero angular momentum}.
\newblock \emph{\bibinfo{journal}{Phys. Rev. B}} \textbf{\bibinfo{volume}{62}},
  \bibinfo{pages}{4880--4889} (\bibinfo{year}{2000}).

\bibitem{lin2021complex}
\bibinfo{author}{Lin, Y.-P.} \& \bibinfo{author}{Nandkishore, R.~M.}
\newblock \bibinfo{title}{{Complex charge density waves at Van Hove singularity
  on hexagonal lattices: Haldane-model phase diagram and potential realization
  in the kagome metals $A$V$_3$Sb$_5$ (A= K, Rb, Cs)}}.
\newblock \emph{\bibinfo{journal}{Physical Review B}}
  \textbf{\bibinfo{volume}{104}}, \bibinfo{pages}{045122}
  (\bibinfo{year}{2021}).

\bibitem{wu2023sublattice}
\bibinfo{author}{Wu, Y.-M.}, \bibinfo{author}{Thomale, R.} \&
  \bibinfo{author}{Raghu, S.}
\newblock \bibinfo{title}{Sublattice interference promotes pair density wave
  order in kagome metals}.
\newblock \emph{\bibinfo{journal}{Physical Review B}}
  \textbf{\bibinfo{volume}{108}}, \bibinfo{pages}{L081117}
  (\bibinfo{year}{2023}).

\bibitem{PhysRevB.80.113102}
\bibinfo{author}{Guo, H.-M.} \& \bibinfo{author}{Franz, M.}
\newblock \bibinfo{title}{Topological insulator on the kagome lattice}.
\newblock \emph{\bibinfo{journal}{Phys. Rev. B}} \textbf{\bibinfo{volume}{80}},
  \bibinfo{pages}{113102} (\bibinfo{year}{2009}).

\bibitem{PhysRevB.85.144402}
\bibinfo{author}{Yu, S.-L.} \& \bibinfo{author}{Li, J.-X.}
\newblock \bibinfo{title}{Chiral superconducting phase and chiral
  spin-density-wave phase in a hubbard model on the kagome lattice}.
\newblock \emph{\bibinfo{journal}{Phys. Rev. B}} \textbf{\bibinfo{volume}{85}},
  \bibinfo{pages}{144402} (\bibinfo{year}{2012}).

\bibitem{PhysRevLett.127.177001}
\bibinfo{author}{Wu, X.} \emph{et~al.}
\newblock \bibinfo{title}{{Nature of Unconventional Pairing in the Kagome
  Superconductors AV$_3$Sb$_5$ (A=K,Rb,Cs)}}.
\newblock \emph{\bibinfo{journal}{Phys. Rev. Lett.}}
  \textbf{\bibinfo{volume}{127}}, \bibinfo{pages}{177001}
  (\bibinfo{year}{2021}).

\bibitem{PhysRevB.106.L060507}
\bibinfo{author}{Lin, Y.-P.} \& \bibinfo{author}{Nandkishore, R.~M.}
\newblock \bibinfo{title}{Multidome superconductivity in charge density wave
  kagome metals}.
\newblock \emph{\bibinfo{journal}{Phys. Rev. B}}
  \textbf{\bibinfo{volume}{106}}, \bibinfo{pages}{L060507}
  (\bibinfo{year}{2022}).

\bibitem{kang2020topological}
\bibinfo{author}{Kang, M.} \emph{et~al.}
\newblock \bibinfo{title}{Topological flat bands in frustrated kagome lattice
  cosn}.
\newblock \emph{\bibinfo{journal}{Nature communications}}
  \textbf{\bibinfo{volume}{11}}, \bibinfo{pages}{4004} (\bibinfo{year}{2020}).

\bibitem{ye2021flat}
\bibinfo{author}{Ye, L.} \emph{et~al.}
\newblock \bibinfo{title}{A flat band-induced correlated kagome metal}.
\newblock \emph{\bibinfo{journal}{arXiv:2106.10824}}  (\bibinfo{year}{2021}).

\bibitem{PhysRevLett.128.096601}
\bibinfo{author}{Huang, H.} \emph{et~al.}
\newblock \bibinfo{title}{Flat-band-induced anomalous anisotropic charge
  transport and orbital magnetism in kagome metal cosn}.
\newblock \emph{\bibinfo{journal}{Phys. Rev. Lett.}}
  \textbf{\bibinfo{volume}{128}}, \bibinfo{pages}{096601}
  (\bibinfo{year}{2022}).

\bibitem{PhysRevLett.101.156402}
\bibinfo{author}{Martin, I.} \& \bibinfo{author}{Batista, C.~D.}
\newblock \bibinfo{title}{Itinerant electron-driven chiral magnetic ordering
  and spontaneous quantum hall effect in triangular lattice models}.
\newblock \emph{\bibinfo{journal}{Phys. Rev. Lett.}}
  \textbf{\bibinfo{volume}{101}}, \bibinfo{pages}{156402}
  (\bibinfo{year}{2008}).

\bibitem{Nandkishore2012}
\bibinfo{author}{Nandkishore, R.}, \bibinfo{author}{Levitov, L.~S.} \&
  \bibinfo{author}{Chubukov, A.~V.}
\newblock \bibinfo{title}{Chiral superconductivity from repulsive interactions
  in doped graphene}.
\newblock \emph{\bibinfo{journal}{Nature Physics}}
  \textbf{\bibinfo{volume}{8}}, \bibinfo{pages}{158--163}
  (\bibinfo{year}{2012}).

\bibitem{ortiz2019new}
\bibinfo{author}{Ortiz, B.~R.} \emph{et~al.}
\newblock \bibinfo{title}{{New kagome prototype materials: discovery of
  KV$_3$Sb$_5$, RbV$_3$Sb$_5$, and CsV$_3$Sb$_5$}}.
\newblock \emph{\bibinfo{journal}{Physical Review Materials}}
  \textbf{\bibinfo{volume}{3}}, \bibinfo{pages}{094407} (\bibinfo{year}{2019}).

\bibitem{kenney2021absence}
\bibinfo{author}{Kenney, E.~M.}, \bibinfo{author}{Ortiz, B.~R.},
  \bibinfo{author}{Wang, C.}, \bibinfo{author}{Wilson, S.~D.} \&
  \bibinfo{author}{Graf, M.~J.}
\newblock \bibinfo{title}{{Absence of local moments in the kagome metal
  KV$_3$Sb$_5$ as determined by muon spin spectroscopy}}.
\newblock \emph{\bibinfo{journal}{Journal of Physics: Condensed Matter}}
  \textbf{\bibinfo{volume}{33}}, \bibinfo{pages}{235801}
  (\bibinfo{year}{2021}).

\bibitem{PhysRevLett.125.247002}
\bibinfo{author}{Ortiz, B.~R.} \emph{et~al.}
\newblock \bibinfo{title}{{CsV$_3$Sb$_5$: A Z2 Topological Kagome Metal with a
  Superconducting Ground State}}.
\newblock \emph{\bibinfo{journal}{Phys. Rev. Lett.}}
  \textbf{\bibinfo{volume}{125}}, \bibinfo{pages}{247002}
  (\bibinfo{year}{2020}).

\bibitem{PhysRevB.104.165136}
\bibinfo{author}{Feng, X.}, \bibinfo{author}{Zhang, Y.},
  \bibinfo{author}{Jiang, K.} \& \bibinfo{author}{Hu, J.}
\newblock \bibinfo{title}{Low-energy effective theory and symmetry
  classification of flux phases on the kagome lattice}.
\newblock \emph{\bibinfo{journal}{Phys. Rev. B}}
  \textbf{\bibinfo{volume}{104}}, \bibinfo{pages}{165136}
  (\bibinfo{year}{2021}).

\bibitem{PhysRevLett.127.217601}
\bibinfo{author}{Denner, M.~M.}, \bibinfo{author}{Thomale, R.} \&
  \bibinfo{author}{Neupert, T.}
\newblock \bibinfo{title}{{Analysis of Charge Order in the Kagome Metal
  AV$_3$Sb$_5$ (A=K,Rb,Cs)}}.
\newblock \emph{\bibinfo{journal}{Phys. Rev. Lett.}}
  \textbf{\bibinfo{volume}{127}}, \bibinfo{pages}{217601}
  (\bibinfo{year}{2021}).

\bibitem{PhysRevB.107.155131}
\bibinfo{author}{Grandi, F.}, \bibinfo{author}{Consiglio, A.},
  \bibinfo{author}{Sentef, M.~A.}, \bibinfo{author}{Thomale, R.} \&
  \bibinfo{author}{Kennes, D.~M.}
\newblock \bibinfo{title}{Theory of nematic charge orders in kagome metals}.
\newblock \emph{\bibinfo{journal}{Phys. Rev. B}}
  \textbf{\bibinfo{volume}{107}}, \bibinfo{pages}{155131}
  (\bibinfo{year}{2023}).

\bibitem{chen2021roton}
\bibinfo{author}{Chen, H.} \emph{et~al.}
\newblock \bibinfo{title}{Roton pair density wave in a strong-coupling kagome
  superconductor}.
\newblock \emph{\bibinfo{journal}{Nature}} \textbf{\bibinfo{volume}{599}},
  \bibinfo{pages}{222--228} (\bibinfo{year}{2021}).

\bibitem{CoSn_yu2011near}
\bibinfo{author}{Yu, X.} \emph{et~al.}
\newblock \bibinfo{title}{{Near room-temperature formation of a skyrmion
  crystal in thin-films of the helimagnet FeGe}}.
\newblock \emph{\bibinfo{journal}{Nature materials}}
  \textbf{\bibinfo{volume}{10}}, \bibinfo{pages}{106--109}
  (\bibinfo{year}{2011}).

\bibitem{CoSn_xie2021spin}
\bibinfo{author}{Xie, Y.} \emph{et~al.}
\newblock \bibinfo{title}{{Spin excitations in metallic kagome lattice FeSn and
  CoSn}}.
\newblock \emph{\bibinfo{journal}{Communications Physics}}
  \textbf{\bibinfo{volume}{4}}, \bibinfo{pages}{240} (\bibinfo{year}{2021}).

\bibitem{CoSn_bak1980theory}
\bibinfo{author}{Bak, P.} \& \bibinfo{author}{Jensen, M.~H.}
\newblock \bibinfo{title}{{Theory of helical magnetic structures and phase
  transitions in MnSi and FeGe}}.
\newblock \emph{\bibinfo{journal}{Journal of Physics C: Solid State Physics}}
  \textbf{\bibinfo{volume}{13}}, \bibinfo{pages}{L881} (\bibinfo{year}{1980}).

\bibitem{CoSn_kang2020dirac}
\bibinfo{author}{Kang, M.} \emph{et~al.}
\newblock \bibinfo{title}{{Dirac fermions and flat bands in the ideal kagome
  metal FeSn}}.
\newblock \emph{\bibinfo{journal}{Nature materials}}
  \textbf{\bibinfo{volume}{19}}, \bibinfo{pages}{163--169}
  (\bibinfo{year}{2020}).

\bibitem{CoSn_meier2020flat}
\bibinfo{author}{Meier, W.~R.} \emph{et~al.}
\newblock \bibinfo{title}{{Flat bands in the CoSn-type compounds}}.
\newblock \emph{\bibinfo{journal}{Physical Review B}}
  \textbf{\bibinfo{volume}{102}}, \bibinfo{pages}{075148}
  (\bibinfo{year}{2020}).

\bibitem{CoSn_liu2020orbital}
\bibinfo{author}{Liu, Z.} \emph{et~al.}
\newblock \bibinfo{title}{{Orbital-selective Dirac fermions and extremely flat
  bands in frustrated kagome-lattice metal CoSn}}.
\newblock \emph{\bibinfo{journal}{Nature communications}}
  \textbf{\bibinfo{volume}{11}}, \bibinfo{pages}{4002} (\bibinfo{year}{2020}).

\bibitem{CoSn_sales2021tuning}
\bibinfo{author}{Sales, B.} \emph{et~al.}
\newblock \bibinfo{title}{{Tuning the flat bands of the kagome metal CoSn with
  Fe, In, or Ni doping}}.
\newblock \emph{\bibinfo{journal}{Physical Review Materials}}
  \textbf{\bibinfo{volume}{5}}, \bibinfo{pages}{044202} (\bibinfo{year}{2021}).

\bibitem{CoSn_teng2023magnetism}
\bibinfo{author}{Teng, X.} \emph{et~al.}
\newblock \bibinfo{title}{{Magnetism and charge density wave order in kagome
  FeGe}}.
\newblock \emph{\bibinfo{journal}{Nature Physics}} \bibinfo{pages}{1--9}
  (\bibinfo{year}{2023}).

\bibitem{CoSn_kang2020topological}
\bibinfo{author}{Kang, M.} \emph{et~al.}
\newblock \bibinfo{title}{{Topological flat bands in frustrated kagome lattice
  CoSn}}.
\newblock \emph{\bibinfo{journal}{Nature communications}}
  \textbf{\bibinfo{volume}{11}}, \bibinfo{pages}{4004} (\bibinfo{year}{2020}).

\bibitem{CoSn_sales2019electronic}
\bibinfo{author}{Sales, B.~C.} \emph{et~al.}
\newblock \bibinfo{title}{{Electronic, magnetic, and thermodynamic properties
  of the kagome layer compound FeSn}}.
\newblock \emph{\bibinfo{journal}{Physical Review Materials}}
  \textbf{\bibinfo{volume}{3}}, \bibinfo{pages}{114203} (\bibinfo{year}{2019}).

\bibitem{166_arachchige2022charge}
\bibinfo{author}{Arachchige, H. W.~S.} \emph{et~al.}
\newblock \bibinfo{title}{{Charge Density Wave in Kagome Lattice Intermetallic
  ScV$_6$Sn$_6$}}.
\newblock \emph{\bibinfo{journal}{Physical Review Letters}}
  \textbf{\bibinfo{volume}{129}}, \bibinfo{pages}{216402}
  (\bibinfo{year}{2022}).

\bibitem{166_el1991magnetic}
\bibinfo{author}{El~Idrissi, B.~C.}, \bibinfo{author}{Venturini, G.},
  \bibinfo{author}{Malaman, B.} \& \bibinfo{author}{Fruchart, D.}
\newblock \bibinfo{title}{{Magnetic structures of TbMn$_6$Sn$_6$ and
  HoMn$_6$Sn$_6$ compounds from neutron diffraction study}}.
\newblock \emph{\bibinfo{journal}{Journal of the Less Common Metals}}
  \textbf{\bibinfo{volume}{175}}, \bibinfo{pages}{143--154}
  (\bibinfo{year}{1991}).

\bibitem{166_ghimire2020competing}
\bibinfo{author}{Ghimire, N.~J.} \emph{et~al.}
\newblock \bibinfo{title}{{Competing magnetic phases and fluctuation-driven
  scalar spin chirality in the kagome metal YMn$_6$Sn$_6$}}.
\newblock \emph{\bibinfo{journal}{Science Advances}}
  \textbf{\bibinfo{volume}{6}}, \bibinfo{pages}{eabe2680}
  (\bibinfo{year}{2020}).

\bibitem{166_lee2022anisotropic}
\bibinfo{author}{Lee, J.} \& \bibinfo{author}{Mun, E.}
\newblock \bibinfo{title}{{Anisotropic magnetic property of single crystals
  RV$_6$Sn$_6$ (R= Y, Gd- Tm, Lu)}}.
\newblock \emph{\bibinfo{journal}{Physical Review Materials}}
  \textbf{\bibinfo{volume}{6}}, \bibinfo{pages}{083401} (\bibinfo{year}{2022}).

\bibitem{166_peng2021realizing}
\bibinfo{author}{Peng, S.} \emph{et~al.}
\newblock \bibinfo{title}{{Realizing Kagome Band Structure in Two-Dimensional
  Kagome Surface States of RV$_6$Sn$_6$ (R= Gd, Ho)}}.
\newblock \emph{\bibinfo{journal}{Physical review letters}}
  \textbf{\bibinfo{volume}{127}}, \bibinfo{pages}{266401}
  (\bibinfo{year}{2021}).

\bibitem{166_pokharel2021electronic}
\bibinfo{author}{Pokharel, G.} \emph{et~al.}
\newblock \bibinfo{title}{{Electronic properties of the topological kagome
  metals YV$_6$Sn$_6$ and GdV$_6$Sn$_6$}}.
\newblock \emph{\bibinfo{journal}{Physical Review B}}
  \textbf{\bibinfo{volume}{104}}, \bibinfo{pages}{235139}
  (\bibinfo{year}{2021}).

\bibitem{166_pokharel2022highly}
\bibinfo{author}{Pokharel, G.} \emph{et~al.}
\newblock \bibinfo{title}{{Highly anisotropic magnetism in the vanadium-based
  kagome metal TbV$_6$Sn$_6$}}.
\newblock \emph{\bibinfo{journal}{Physical Review Materials}}
  \textbf{\bibinfo{volume}{6}}, \bibinfo{pages}{104202} (\bibinfo{year}{2022}).

\bibitem{166_wang2021field}
\bibinfo{author}{Wang, Q.} \emph{et~al.}
\newblock \bibinfo{title}{{Field-induced topological Hall effect and double-fan
  spin structure with a c-axis component in the metallic kagome
  antiferromagnetic compound YMn$_6$Sn$_6$}}.
\newblock \emph{\bibinfo{journal}{Physical Review B}}
  \textbf{\bibinfo{volume}{103}}, \bibinfo{pages}{014416}
  (\bibinfo{year}{2021}).

\bibitem{166_yin2020quantum}
\bibinfo{author}{Yin, J.-X.} \emph{et~al.}
\newblock \bibinfo{title}{{Quantum-limit Chern topological magnetism in
  TbMn$_6$Sn$_6$}}.
\newblock \emph{\bibinfo{journal}{Nature}} \textbf{\bibinfo{volume}{583}},
  \bibinfo{pages}{533--536} (\bibinfo{year}{2020}).

\bibitem{166_zhang2022electronic}
\bibinfo{author}{Zhang, X.} \emph{et~al.}
\newblock \bibinfo{title}{{Electronic and magnetic properties of intermetallic
  kagome magnets RV$_6$Sn$_6$ (R= Tb- Tm)}}.
\newblock \emph{\bibinfo{journal}{Physical Review Materials}}
  \textbf{\bibinfo{volume}{6}}, \bibinfo{pages}{105001} (\bibinfo{year}{2022}).

\bibitem{thinsong2021competing}
\bibinfo{author}{Song, B.} \emph{et~al.}
\newblock \bibinfo{title}{Competing superconductivity and charge-density wave
  in kagome metal : evidence from their evolutions with sample thickness}.
\newblock \emph{\bibinfo{journal}{arXiv:2105.09248}}  (\bibinfo{year}{2021}).

\bibitem{thinwei2022linear}
\bibinfo{author}{Wei, X.} \emph{et~al.}
\newblock \bibinfo{title}{Linear nonsaturating magnetoresistance in kagome
  superconductor thin flakes}.
\newblock \emph{\bibinfo{journal}{2D Materials}} \textbf{\bibinfo{volume}{10}},
  \bibinfo{pages}{015010} (\bibinfo{year}{2022}).

\bibitem{thinwu2022nonreciprocal}
\bibinfo{author}{Wu, Y.} \emph{et~al.}
\newblock \bibinfo{title}{{Nonreciprocal charge transport in topological kagome
  superconductor CsV$_3$Sb$_5$}}.
\newblock \emph{\bibinfo{journal}{npj Quantum Materials}}
  \textbf{\bibinfo{volume}{7}}, \bibinfo{pages}{105} (\bibinfo{year}{2022}).

\bibitem{thinsong2021competition}
\bibinfo{author}{Song, Y.} \emph{et~al.}
\newblock \bibinfo{title}{{Competition of superconductivity and charge density
  wave in selective oxidized CsV$_3$Sb$_5$ thin flakes}}.
\newblock \emph{\bibinfo{journal}{Physical review letters}}
  \textbf{\bibinfo{volume}{127}}, \bibinfo{pages}{237001}
  (\bibinfo{year}{2021}).

\bibitem{klepp1996crystal}
\bibinfo{author}{Klepp, K.} \& \bibinfo{author}{Weithaler, C.}
\newblock \bibinfo{title}{{The crystal structures of CsAu$_3$S$_2$,
  RbAu$_3$Se$_2$ and CsAu$_3$Se$_2$ and their relationship to the CsCu$_3$S$_2$
  structure type}}.
\newblock \emph{\bibinfo{journal}{Journal of alloys and compounds}}
  \textbf{\bibinfo{volume}{243}}, \bibinfo{pages}{1--5} (\bibinfo{year}{1996}).

\bibitem{burschka1980cscu}
\bibinfo{author}{Burschka, C.}
\newblock \bibinfo{title}{{CsCu$_4$S$_3$ und CsCu$_3$S$_2$: Sulfide mit
  tetraedrisch und linear koordiniertem Kupfer}}.
\newblock \emph{\bibinfo{journal}{Z. anorg. allg. Ohem.}}
  \textbf{\bibinfo{volume}{483}}, \bibinfo{pages}{65--71}
  (\bibinfo{year}{1980}).

\bibitem{savelsberg1978darstellung}
\bibinfo{author}{Savelsberg, G.} \& \bibinfo{author}{SCHAFER, H.}
\newblock \bibinfo{title}{{Darstellung und Kristallstruktur von
  K$_3$CU$_3$P$_2$}}.
\newblock \emph{\bibinfo{journal}{Z. Naturforsch. B}}
  \textbf{\bibinfo{volume}{33}}, \bibinfo{pages}{590--592}
  (\bibinfo{year}{1978}).

\bibitem{bronger1971cs2pd3s4}
\bibinfo{author}{Bronger, W.} \& \bibinfo{author}{Huster, J.}
\newblock \bibinfo{title}{{Cs$_2$Pd$_3$S$_4$, ein neuer schichtenstrukturtyp}}.
\newblock \emph{\bibinfo{journal}{Journal of the Less Common Metals}}
  \textbf{\bibinfo{volume}{23}}, \bibinfo{pages}{67--72}
  (\bibinfo{year}{1971}).

\bibitem{PhysRevMaterials.7.064201}
\bibinfo{author}{Ortiz, B.~R.} \emph{et~al.}
\newblock \bibinfo{title}{{YbV$_3$Sb$_4$ and EuV$_3$Sb$_4$ vanadium-based
  kagome metals with Yb$^{2+}$ and Eu$^{2+}$ zigzag chains}}.
\newblock \emph{\bibinfo{journal}{Phys. Rev. Mater.}}
  \textbf{\bibinfo{volume}{7}}, \bibinfo{pages}{064201} (\bibinfo{year}{2023}).

\bibitem{ortiz2023evolution}
\bibinfo{author}{Ortiz, B.~R.} \emph{et~al.}
\newblock \bibinfo{title}{{Evolution of highly anisotropic magnetism in the
  titanium-based kagome metals LnTi$_3$Bi$_4$ (Ln: La...Gd$^{3+}$, Eu$^{2+}$,
  Yb$^{2+}$}}.
\newblock \emph{\bibinfo{journal}{arXiv:2308.16138}}  (\bibinfo{year}{2023}).

\bibitem{ovchinnikov2018synthesis}
\bibinfo{author}{Ovchinnikov, A.} \& \bibinfo{author}{Bobev, S.}
\newblock \bibinfo{title}{{Synthesis, Crystal and Electronic Structure of the
  Titanium Bismuthides Sr$_5$Ti$_{12}$Bi$_{19+x}$, Ba$_5$Ti$_{12}$Bi$_{19+x}$,
  and Sr$_{5-\delta}$Eu$_\delta$Ti$_{12}$Bi$_{19+x}$ (x=0.5--1.0; $\delta$=2.4,
  4.0)}}.
\newblock \emph{\bibinfo{journal}{Eur. J. Inorg. Chem.}}
  \textbf{\bibinfo{volume}{2018}}, \bibinfo{pages}{1266--1274}
  (\bibinfo{year}{2018}).

\bibitem{ovchinnikov2019bismuth}
\bibinfo{author}{Ovchinnikov, A.} \& \bibinfo{author}{Bobev, S.}
\newblock \bibinfo{title}{Bismuth as a reactive solvent in the synthesis of
  multicomponent transition-metal-bearing bismuthides}.
\newblock \emph{\bibinfo{journal}{Inorg. Chem.}} \textbf{\bibinfo{volume}{59}},
  \bibinfo{pages}{3459--3470} (\bibinfo{year}{2019}).

\bibitem{jovanovic2022simple}
\bibinfo{author}{Jovanovic, M.} \& \bibinfo{author}{Schoop, L.~M.}
\newblock \bibinfo{title}{Simple chemical rules for predicting band structures
  of kagome materials}.
\newblock \emph{\bibinfo{journal}{Journal of the American Chemical Society}}
  \textbf{\bibinfo{volume}{144}}, \bibinfo{pages}{10978--10991}
  (\bibinfo{year}{2022}).

\bibitem{PhysRevLett.127.046401}
\bibinfo{author}{Tan, H.}, \bibinfo{author}{Liu, Y.}, \bibinfo{author}{Wang,
  Z.} \& \bibinfo{author}{Yan, B.}
\newblock \bibinfo{title}{Charge density waves and electronic properties of
  superconducting kagome metals}.
\newblock \emph{\bibinfo{journal}{Phys. Rev. Lett.}}
  \textbf{\bibinfo{volume}{127}}, \bibinfo{pages}{046401}
  (\bibinfo{year}{2021}).

\bibitem{kang2022twofold}
\bibinfo{author}{Kang, M.} \emph{et~al.}
\newblock \bibinfo{title}{{Twofold van Hove singularity and origin of charge
  order in topological kagome superconductor CsV$_3$Sb$_5$}}.
\newblock \emph{\bibinfo{journal}{Nature Physics}}
  \textbf{\bibinfo{volume}{18}}, \bibinfo{pages}{301--308}
  (\bibinfo{year}{2022}).

\bibitem{hu2022rich}
\bibinfo{author}{Hu, Y.} \emph{et~al.}
\newblock \bibinfo{title}{{Rich nature of Van Hove singularities in Kagome
  superconductor CsV$_3$Sb$_5$}}.
\newblock \emph{\bibinfo{journal}{Nature Communications}}
  \textbf{\bibinfo{volume}{13}}, \bibinfo{pages}{2220} (\bibinfo{year}{2022}).

\bibitem{Luo2022}
\bibinfo{author}{Luo, H.} \emph{et~al.}
\newblock \bibinfo{title}{{Electronic nature of charge density wave and
  electron-phonon coupling in kagome superconductor KV$_3$Sb$_5$}}.
\newblock \emph{\bibinfo{journal}{Nature Communications}}
  \textbf{\bibinfo{volume}{13}}, \bibinfo{pages}{273} (\bibinfo{year}{2022}).

\bibitem{kaboudvand2022fermi}
\bibinfo{author}{Kaboudvand, F.}, \bibinfo{author}{Teicher, S.~M.},
  \bibinfo{author}{Wilson, S.~D.}, \bibinfo{author}{Seshadri, R.} \&
  \bibinfo{author}{Johannes, M.~D.}
\newblock \bibinfo{title}{{Fermi surface nesting and the Lindhard response
  function in the kagome superconductor CsV$_3$Sb$_5$}}.
\newblock \emph{\bibinfo{journal}{Applied Physics Letters}}
  \textbf{\bibinfo{volume}{120}} (\bibinfo{year}{2022}).

\bibitem{PhysRevB.105.L140501}
\bibinfo{author}{Xie, Y.} \emph{et~al.}
\newblock \bibinfo{title}{{Electron-phonon coupling in the charge density wave
  state of CsV$_3$Sb$_5$}}.
\newblock \emph{\bibinfo{journal}{Phys. Rev. B}}
  \textbf{\bibinfo{volume}{105}}, \bibinfo{pages}{L140501}
  (\bibinfo{year}{2022}).

\bibitem{PhysRevB.106.L081107}
\bibinfo{author}{Ferrari, F.}, \bibinfo{author}{Becca, F.} \&
  \bibinfo{author}{Valent\'{\i}, R.}
\newblock \bibinfo{title}{{Charge density waves in kagome-lattice extended
  Hubbard models at the Van Hove filling}}.
\newblock \emph{\bibinfo{journal}{Phys. Rev. B}}
  \textbf{\bibinfo{volume}{106}}, \bibinfo{pages}{L081107}
  (\bibinfo{year}{2022}).

\bibitem{ece2022optical}
\bibinfo{author}{Ece, U.}, \bibinfo{author}{Ortiz, B.~R.},
  \bibinfo{author}{Wilson, S.~D.}, \bibinfo{author}{Dressel, M.} \&
  \bibinfo{author}{Tsirlin, A.~A.}
\newblock \bibinfo{title}{{Optical detection of the density-wave instability in
  the kagome metal KV$_3$Sb$_5$}}.
\newblock \emph{\bibinfo{journal}{NPJ Quantum Materials}}
  \textbf{\bibinfo{volume}{7}} (\bibinfo{year}{2022}).

\bibitem{PhysRevB.104.045130}
\bibinfo{author}{Uykur, E.} \emph{et~al.}
\newblock \bibinfo{title}{{Low-energy optical properties of the nonmagnetic
  kagome metal CsV$_3$Sb$_5$}}.
\newblock \emph{\bibinfo{journal}{Phys. Rev. B}}
  \textbf{\bibinfo{volume}{104}}, \bibinfo{pages}{045130}
  (\bibinfo{year}{2021}).

\bibitem{PhysRevB.105.245123}
\bibinfo{author}{Wenzel, M.} \emph{et~al.}
\newblock \bibinfo{title}{{Optical study of RbV$_3$Sb$_5$: Multiple
  density-wave gaps and phonon anomalies}}.
\newblock \emph{\bibinfo{journal}{Phys. Rev. B}}
  \textbf{\bibinfo{volume}{105}}, \bibinfo{pages}{245123}
  (\bibinfo{year}{2022}).

\bibitem{PhysRevX.11.031026}
\bibinfo{author}{Liang, Z.} \emph{et~al.}
\newblock \bibinfo{title}{{Three-Dimensional Charge Density Wave and
  Surface-Dependent Vortex-Core States in a Kagome Superconductor
  CsV$_3$Sb$_5$}}.
\newblock \emph{\bibinfo{journal}{Phys. Rev. X}} \textbf{\bibinfo{volume}{11}},
  \bibinfo{pages}{031026} (\bibinfo{year}{2021}).

\bibitem{Huai_2022}
\bibinfo{author}{Huai, L.} \emph{et~al.}
\newblock \bibinfo{title}{{Surface-induced orbital-selective band
  reconstruction in kagome superconductor CsV$_3$Sb$_5$}}.
\newblock \emph{\bibinfo{journal}{Chinese Physics B}}
  \textbf{\bibinfo{volume}{31}}, \bibinfo{pages}{057403}
  (\bibinfo{year}{2022}).
\newblock \urlprefix\url{https://dx.doi.org/10.1088/1674-1056/ac4f50}.

\bibitem{PhysRevB.104.205129}
\bibinfo{author}{LaBollita, H.} \& \bibinfo{author}{Botana, A.~S.}
\newblock \bibinfo{title}{{Tuning the Van Hove singularities in AV$_3$Sb$_5$
  (A=K,Rb,Cs) via pressure and doping}}.
\newblock \emph{\bibinfo{journal}{Phys. Rev. B}}
  \textbf{\bibinfo{volume}{104}}, \bibinfo{pages}{205129}
  (\bibinfo{year}{2021}).

\bibitem{PhysRevMaterials.7.024806}
\bibinfo{author}{Kautzsch, L.} \emph{et~al.}
\newblock \bibinfo{title}{{Structural evolution of the kagome superconductors
  AV$_3$Sb$_5$ (A = K, Rb, and Cs) through charge density wave order}}.
\newblock \emph{\bibinfo{journal}{Phys. Rev. Mater.}}
  \textbf{\bibinfo{volume}{7}}, \bibinfo{pages}{024806} (\bibinfo{year}{2023}).

\bibitem{PhysRevB.105.235145}
\bibinfo{author}{Jeong, M.~Y.} \emph{et~al.}
\newblock \bibinfo{title}{{Crucial role of out-of-plane Sb $p$ orbitals in Van
  Hove singularity formation and electronic correlations in the superconducting
  kagome metal CsV$_3$Sb$_5$}}.
\newblock \emph{\bibinfo{journal}{Phys. Rev. B}}
  \textbf{\bibinfo{volume}{105}}, \bibinfo{pages}{235145}
  (\bibinfo{year}{2022}).

\bibitem{PhysRevB.107.205131}
\bibinfo{author}{Ritz, E.~T.}, \bibinfo{author}{Fernandes, R.~M.} \&
  \bibinfo{author}{Birol, T.}
\newblock \bibinfo{title}{{Impact of Sb degrees of freedom on the charge
  density wave phase diagram of the kagome metal CsV$_3$Sb$_5$}}.
\newblock \emph{\bibinfo{journal}{Phys. Rev. B}}
  \textbf{\bibinfo{volume}{107}}, \bibinfo{pages}{205131}
  (\bibinfo{year}{2023}).

\bibitem{PhysRevB.108.075102}
\bibinfo{author}{Li, H.}, \bibinfo{author}{Liu, X.}, \bibinfo{author}{Kim,
  Y.~B.} \& \bibinfo{author}{Kee, H.-Y.}
\newblock \bibinfo{title}{{Origin of $\pi$-shifted three-dimensional charge
  density waves in the kagom\'e metal AV$_3$Sb$_5$ (A=Cs, Rb, K)}}.
\newblock \emph{\bibinfo{journal}{Phys. Rev. B}}
  \textbf{\bibinfo{volume}{108}}, \bibinfo{pages}{075102}
  (\bibinfo{year}{2023}).

\bibitem{PhysRevX.11.041030}
\bibinfo{author}{Ortiz, B.~R.} \emph{et~al.}
\newblock \bibinfo{title}{{Fermi Surface Mapping and the Nature of
  Charge-Density-Wave Order in the Kagome Superconductor CsV$_3$Sb$_5$}}.
\newblock \emph{\bibinfo{journal}{Phys. Rev. X}} \textbf{\bibinfo{volume}{11}},
  \bibinfo{pages}{041030} (\bibinfo{year}{2021}).

\bibitem{PhysRevLett.127.207002}
\bibinfo{author}{Fu, Y.} \emph{et~al.}
\newblock \bibinfo{title}{{Quantum Transport Evidence of Topological Band
  Structures of Kagome Superconductor CsV$_3$Sb$_5$}}.
\newblock \emph{\bibinfo{journal}{Phys. Rev. Lett.}}
  \textbf{\bibinfo{volume}{127}}, \bibinfo{pages}{207002}
  (\bibinfo{year}{2021}).

\bibitem{PhysRevB.105.024508}
\bibinfo{author}{Shrestha, K.} \emph{et~al.}
\newblock \bibinfo{title}{{Nontrivial Fermi surface topology of the kagome
  superconductor CsV$_3$Sb$_5$ probed by de Haas--van Alphen oscillations}}.
\newblock \emph{\bibinfo{journal}{Phys. Rev. B}}
  \textbf{\bibinfo{volume}{105}}, \bibinfo{pages}{024508}
  (\bibinfo{year}{2022}).

\bibitem{PhysRevB.107.155128}
\bibinfo{author}{Shrestha, K.} \emph{et~al.}
\newblock \bibinfo{title}{{High quantum oscillation frequencies and nontrivial
  topology in kagome superconductor KV$_3$Sb$_5$ probed by torque magnetometry
  up to 45 T}}.
\newblock \emph{\bibinfo{journal}{Phys. Rev. B}}
  \textbf{\bibinfo{volume}{107}}, \bibinfo{pages}{155128}
  (\bibinfo{year}{2023}).

\bibitem{PhysRevB.107.075120}
\bibinfo{author}{Shrestha, K.} \emph{et~al.}
\newblock \bibinfo{title}{{Fermi surface mapping of the kagome superconductor
  RbV$_3$Sb$_5$ using de Haas-van Alphen oscillations}}.
\newblock \emph{\bibinfo{journal}{Phys. Rev. B}}
  \textbf{\bibinfo{volume}{107}}, \bibinfo{pages}{075120}
  (\bibinfo{year}{2023}).

\bibitem{PhysRevLett.129.157001}
\bibinfo{author}{Broyles, C.} \emph{et~al.}
\newblock \bibinfo{title}{{Effect of the Interlayer Ordering on the Fermi
  Surface of Kagome Superconductor CsV$_3$Sb$_5$ Revealed by Quantum
  Oscillations}}.
\newblock \emph{\bibinfo{journal}{Phys. Rev. Lett.}}
  \textbf{\bibinfo{volume}{129}}, \bibinfo{pages}{157001}
  (\bibinfo{year}{2022}).

\bibitem{hu2022topological}
\bibinfo{author}{Hu, Y.} \emph{et~al.}
\newblock \bibinfo{title}{{Topological surface states and flat bands in the
  kagome superconductor CsV$_3$Sb$_5$}}.
\newblock \emph{\bibinfo{journal}{Science Bulletin}}
  \textbf{\bibinfo{volume}{67}}, \bibinfo{pages}{495--500}
  (\bibinfo{year}{2022}).

\bibitem{PhysRevLett.100.096407}
\bibinfo{author}{Fu, L.} \& \bibinfo{author}{Kane, C.~L.}
\newblock \bibinfo{title}{Superconducting proximity effect and majorana
  fermions at the surface of a topological insulator}.
\newblock \emph{\bibinfo{journal}{Phys. Rev. Lett.}}
  \textbf{\bibinfo{volume}{100}}, \bibinfo{pages}{096407}
  (\bibinfo{year}{2008}).

\bibitem{PhysRevX.11.031050}
\bibinfo{author}{Li, H.} \emph{et~al.}
\newblock \bibinfo{title}{{Observation of Unconventional Charge Density Wave
  without Acoustic Phonon Anomaly in Kagome Superconductors AV$_3$Sb$_5$ (A=Rb,
  Cs)}}.
\newblock \emph{\bibinfo{journal}{Phys. Rev. X}} \textbf{\bibinfo{volume}{11}},
  \bibinfo{pages}{031050} (\bibinfo{year}{2021}).

\bibitem{Jiang2021}
\bibinfo{author}{Jiang, Y.-X.} \emph{et~al.}
\newblock \bibinfo{title}{{Unconventional chiral charge order in kagome
  superconductor KV$_3$Sb$_5$}}.
\newblock \emph{\bibinfo{journal}{Nature Materials}}
  \textbf{\bibinfo{volume}{20}}, \bibinfo{pages}{1353--1357}
  (\bibinfo{year}{2021}).

\bibitem{Zhao2021}
\bibinfo{author}{Zhao, H.} \emph{et~al.}
\newblock \bibinfo{title}{{Cascade of correlated electron states in the kagome
  superconductor CsV$_3$Sb$_5$}}.
\newblock \emph{\bibinfo{journal}{Nature}} \textbf{\bibinfo{volume}{599}},
  \bibinfo{pages}{216--221} (\bibinfo{year}{2021}).

\bibitem{PhysRevB.104.035131}
\bibinfo{author}{Shumiya, N.} \emph{et~al.}
\newblock \bibinfo{title}{{Intrinsic nature of chiral charge order in the
  kagome superconductor RbV$_3$Sb$_5$}}.
\newblock \emph{\bibinfo{journal}{Phys. Rev. B}}
  \textbf{\bibinfo{volume}{104}}, \bibinfo{pages}{035131}
  (\bibinfo{year}{2021}).

\bibitem{PhysRevB.104.214513}
\bibinfo{author}{Christensen, M.~H.}, \bibinfo{author}{Birol, T.},
  \bibinfo{author}{Andersen, B.~M.} \& \bibinfo{author}{Fernandes, R.~M.}
\newblock \bibinfo{title}{{Theory of the charge density wave in AV$_3$Sb$_5$
  kagome metals}}.
\newblock \emph{\bibinfo{journal}{Phys. Rev. B}}
  \textbf{\bibinfo{volume}{104}}, \bibinfo{pages}{214513}
  (\bibinfo{year}{2021}).

\bibitem{PhysRevMaterials.6.015001}
\bibinfo{author}{Subedi, A.}
\newblock \bibinfo{title}{{Hexagonal-to-base-centered-orthorhombic $4Q$ charge
  density wave order in kagome metals KV$_3$Sb$_5$, RbV$_3$Sb$_5$ and
  CsV$_3$Sb$_5$}}.
\newblock \emph{\bibinfo{journal}{Phys. Rev. Mater.}}
  \textbf{\bibinfo{volume}{6}}, \bibinfo{pages}{015001} (\bibinfo{year}{2022}).

\bibitem{PhysRevResearch.5.L012017}
\bibinfo{author}{Frassineti, J.} \emph{et~al.}
\newblock \bibinfo{title}{{Microscopic nature of the charge-density wave in the
  kagome superconductor RbV$_3$Sb$_5$}}.
\newblock \emph{\bibinfo{journal}{Phys. Rev. Res.}}
  \textbf{\bibinfo{volume}{5}}, \bibinfo{pages}{L012017}
  (\bibinfo{year}{2023}).

\bibitem{Kang2023}
\bibinfo{author}{Kang, M.} \emph{et~al.}
\newblock \bibinfo{title}{Charge order landscape and competition with
  superconductivity in kagome metals}.
\newblock \emph{\bibinfo{journal}{Nature Materials}}
  \textbf{\bibinfo{volume}{22}}, \bibinfo{pages}{186--193}
  (\bibinfo{year}{2023}).

\bibitem{PhysRevB.106.L241106}
\bibinfo{author}{Hu, Y.} \emph{et~al.}
\newblock \bibinfo{title}{{Coexistence of trihexagonal and star-of-David
  pattern in the charge density wave of the kagome superconductor
  AV$_3$Sb$_5$}}.
\newblock \emph{\bibinfo{journal}{Phys. Rev. B}}
  \textbf{\bibinfo{volume}{106}}, \bibinfo{pages}{L241106}
  (\bibinfo{year}{2022}).

\bibitem{PhysRevB.105.195136}
\bibinfo{author}{Stahl, Q.} \emph{et~al.}
\newblock \bibinfo{title}{{Temperature-driven reorganization of electronic
  order in CsV$_3$Sb$_5$}}.
\newblock \emph{\bibinfo{journal}{Phys. Rev. B}}
  \textbf{\bibinfo{volume}{105}}, \bibinfo{pages}{195136}
  (\bibinfo{year}{2022}).

\bibitem{PhysRevResearch.5.L012032}
\bibinfo{author}{Xiao, Q.} \emph{et~al.}
\newblock \bibinfo{title}{{Coexistence of multiple stacking charge density
  waves in kagome superconductor CsV$_3$Sb$_5$}}.
\newblock \emph{\bibinfo{journal}{Phys. Rev. Res.}}
  \textbf{\bibinfo{volume}{5}}, \bibinfo{pages}{L012032}
  (\bibinfo{year}{2023}).

\bibitem{PhysRevB.104.035142}
\bibinfo{author}{Park, T.}, \bibinfo{author}{Ye, M.} \&
  \bibinfo{author}{Balents, L.}
\newblock \bibinfo{title}{Electronic instabilities of kagome metals: Saddle
  points and landau theory}.
\newblock \emph{\bibinfo{journal}{Phys. Rev. B}}
  \textbf{\bibinfo{volume}{104}}, \bibinfo{pages}{035142}
  (\bibinfo{year}{2021}).

\bibitem{PhysRevB.104.075148}
\bibinfo{author}{Wang, Z.} \emph{et~al.}
\newblock \bibinfo{title}{{Electronic nature of chiral charge order in the
  kagome superconductor CsV$_3$Sb$_5$}}.
\newblock \emph{\bibinfo{journal}{Phys. Rev. B}}
  \textbf{\bibinfo{volume}{104}}, \bibinfo{pages}{075148}
  (\bibinfo{year}{2021}).

\bibitem{Li2022_NatPhys}
\bibinfo{author}{Li, H.} \emph{et~al.}
\newblock \bibinfo{title}{{Rotation symmetry breaking in the normal state of a
  kagome superconductor KV$_3$Sb$_5$}}.
\newblock \emph{\bibinfo{journal}{Nature Physics}}
  \textbf{\bibinfo{volume}{18}}, \bibinfo{pages}{265--270}
  (\bibinfo{year}{2022}).

\bibitem{Li2023}
\bibinfo{author}{Li, H.} \emph{et~al.}
\newblock \bibinfo{title}{{Unidirectional coherent quasiparticles in the
  high-temperature rotational symmetry broken phase of AV$_3$Sb$_5$ kagome
  superconductors}}.
\newblock \emph{\bibinfo{journal}{Nature Physics}}
  \textbf{\bibinfo{volume}{19}}, \bibinfo{pages}{637--643}
  (\bibinfo{year}{2023}).

\bibitem{PhysRevB.105.045102}
\bibinfo{author}{Li, H.} \emph{et~al.}
\newblock \bibinfo{title}{{No observation of chiral flux current in the
  topological kagome metal CsV$_3$Sb$_5$}}.
\newblock \emph{\bibinfo{journal}{Phys. Rev. B}}
  \textbf{\bibinfo{volume}{105}}, \bibinfo{pages}{045102}
  (\bibinfo{year}{2022}).

\bibitem{Xu2022}
\bibinfo{author}{Xu, Y.} \emph{et~al.}
\newblock \bibinfo{title}{Three-state nematicity and magneto-optical kerr
  effect in the charge density waves in kagome superconductors}.
\newblock \emph{\bibinfo{journal}{Nature Physics}}
  \textbf{\bibinfo{volume}{18}}, \bibinfo{pages}{1470--1475}
  (\bibinfo{year}{2022}).

\bibitem{PhysRevLett.131.016901}
\bibinfo{author}{Saykin, D.~R.} \emph{et~al.}
\newblock \bibinfo{title}{{High Resolution Polar Kerr Effect Studies of
  CsV$_3$Sb$_5$: Tests for Time-Reversal Symmetry Breaking below the
  Charge-Order Transition}}.
\newblock \emph{\bibinfo{journal}{Phys. Rev. Lett.}}
  \textbf{\bibinfo{volume}{131}}, \bibinfo{pages}{016901}
  (\bibinfo{year}{2023}).

\bibitem{Mielke2022}
\bibinfo{author}{Mielke, C.} \emph{et~al.}
\newblock \bibinfo{title}{Time-reversal symmetry-breaking charge order in a
  kagome superconductor}.
\newblock \emph{\bibinfo{journal}{Nature}} \textbf{\bibinfo{volume}{602}},
  \bibinfo{pages}{245--250} (\bibinfo{year}{2022}).

\bibitem{PhysRevResearch.4.023244}
\bibinfo{author}{Khasanov, R.} \emph{et~al.}
\newblock \bibinfo{title}{{Time-reversal symmetry broken by charge order in
  CsV$_3$Sb$_5$}}.
\newblock \emph{\bibinfo{journal}{Phys. Rev. Res.}}
  \textbf{\bibinfo{volume}{4}}, \bibinfo{pages}{023244} (\bibinfo{year}{2022}).

\bibitem{Guguchia2023}
\bibinfo{author}{Guguchia, Z.} \emph{et~al.}
\newblock \bibinfo{title}{{Tunable unconventional kagome superconductivity in
  charge ordered RbV$_3$Sb$_5$ and KV$_3$Sb$_5$}}.
\newblock \emph{\bibinfo{journal}{Nature Communications}}
  \textbf{\bibinfo{volume}{14}}, \bibinfo{pages}{153} (\bibinfo{year}{2023}).

\bibitem{yu2021evidence}
\bibinfo{author}{Yu, L.} \emph{et~al.}
\newblock \bibinfo{title}{{Evidence of a hidden flux phase in the topological
  kagome metal CsV$_3$Sb$_5$}}.
\newblock \emph{\bibinfo{journal}{arXiv:2107.10714}}  (\bibinfo{year}{2021}).

\bibitem{doi:10.1126/sciadv.abb6003}
\bibinfo{author}{Yang, S.-Y.} \emph{et~al.}
\newblock \bibinfo{title}{{Giant, unconventional anomalous Hall effect in the
  metallic frustrated magnet candidate, KV$_3$Sb$_5$}}.
\newblock \emph{\bibinfo{journal}{Science Advances}}
  \textbf{\bibinfo{volume}{6}}, \bibinfo{pages}{eabb6003}
  (\bibinfo{year}{2020}).
\newblock \eprint{https://www.science.org/doi/pdf/10.1126/sciadv.abb6003}.

\bibitem{Wang_2023_RVS}
\bibinfo{author}{Wang, L.} \emph{et~al.}
\newblock \bibinfo{title}{{Anomalous Hall effect and two-dimensional Fermi
  surfaces in the charge-density-wave state of kagome metal RbV$_3$Sb$_5$}}.
\newblock \emph{\bibinfo{journal}{Journal of Physics: Materials}}
  \textbf{\bibinfo{volume}{6}}, \bibinfo{pages}{02LT01} (\bibinfo{year}{2023}).
\newblock \urlprefix\url{https://dx.doi.org/10.1088/2515-7639/acba46}.

\bibitem{PhysRevB.104.L041103}
\bibinfo{author}{Yu, F.~H.} \emph{et~al.}
\newblock \bibinfo{title}{Concurrence of anomalous hall effect and charge
  density wave in a superconducting topological kagome metal}.
\newblock \emph{\bibinfo{journal}{Phys. Rev. B}}
  \textbf{\bibinfo{volume}{104}}, \bibinfo{pages}{L041103}
  (\bibinfo{year}{2021}).

\bibitem{Guo2022}
\bibinfo{author}{Guo, C.} \emph{et~al.}
\newblock \bibinfo{title}{{Switchable chiral transport in charge-ordered kagome
  metal CsV$_3$Sb$_5$}}.
\newblock \emph{\bibinfo{journal}{Nature}} \textbf{\bibinfo{volume}{611}},
  \bibinfo{pages}{461--466} (\bibinfo{year}{2022}).

\bibitem{asaba2023evidence}
\bibinfo{author}{Asaba, T.} \emph{et~al.}
\newblock \bibinfo{title}{{Evidence for an odd-parity nematic phase above the
  charge density wave transition in kagome metal CsV$_3$Sb$_5$}}
  (\bibinfo{year}{2023}).
\newblock \eprint{2309.16985}.

\bibitem{Subires2023}
\bibinfo{author}{Subires, D.} \emph{et~al.}
\newblock \bibinfo{title}{{Order-disorder charge density wave instability in
  the kagome metal (Cs,Rb)V$_3$Sb$_5$}}.
\newblock \emph{\bibinfo{journal}{Nature Communications}}
  \textbf{\bibinfo{volume}{14}}, \bibinfo{pages}{1015} (\bibinfo{year}{2023}).

\bibitem{PhysRevLett.129.056401}
\bibinfo{author}{Chen, Q.}, \bibinfo{author}{Chen, D.},
  \bibinfo{author}{Schnelle, W.}, \bibinfo{author}{Felser, C.} \&
  \bibinfo{author}{Gaulin, B.~D.}
\newblock \bibinfo{title}{{Charge Density Wave Order and Fluctuations above
  $T_{CDW}$ and below Superconducting $T_c$ in the Kagome Metal
  CsV$_3$Sb$_5$}}.
\newblock \emph{\bibinfo{journal}{Phys. Rev. Lett.}}
  \textbf{\bibinfo{volume}{129}}, \bibinfo{pages}{056401}
  (\bibinfo{year}{2022}).

\bibitem{PhysRevB.105.L201109}
\bibinfo{author}{Chen, D.} \emph{et~al.}
\newblock \bibinfo{title}{{Anomalous thermoelectric effects and quantum
  oscillations in the kagome metal CsV$_3$Sb$_5$}}.
\newblock \emph{\bibinfo{journal}{Phys. Rev. B}}
  \textbf{\bibinfo{volume}{105}}, \bibinfo{pages}{L201109}
  (\bibinfo{year}{2022}).

\bibitem{PhysRevB.105.094507}
\bibinfo{author}{Zhu, C.~C.} \emph{et~al.}
\newblock \bibinfo{title}{{Double-dome superconductivity under pressure in the
  V-based kagome metals AV$_3$Sb$_5$ (A=Rb and K)}}.
\newblock \emph{\bibinfo{journal}{Phys. Rev. B}}
  \textbf{\bibinfo{volume}{105}}, \bibinfo{pages}{094507}
  (\bibinfo{year}{2022}).

\bibitem{PhysRevMaterials.5.L111801}
\bibinfo{author}{Ratcliff, N.}, \bibinfo{author}{Hallett, L.},
  \bibinfo{author}{Ortiz, B.~R.}, \bibinfo{author}{Wilson, S.~D.} \&
  \bibinfo{author}{Harter, J.~W.}
\newblock \bibinfo{title}{{Coherent phonon spectroscopy and interlayer
  modulation of charge density wave order in the kagome metal CsV$_3$Sb$_5$}}.
\newblock \emph{\bibinfo{journal}{Phys. Rev. Mater.}}
  \textbf{\bibinfo{volume}{5}}, \bibinfo{pages}{L111801}
  (\bibinfo{year}{2021}).

\bibitem{PhysRevB.105.155106}
\bibinfo{author}{Wu, S.} \emph{et~al.}
\newblock \bibinfo{title}{{Charge density wave order in the kagome metal
  AV$_3$Sb$_5$ (A=Cs,Rb,K)}}.
\newblock \emph{\bibinfo{journal}{Phys. Rev. B}}
  \textbf{\bibinfo{volume}{105}}, \bibinfo{pages}{155106}
  (\bibinfo{year}{2022}).

\bibitem{PhysRevResearch.4.023215}
\bibinfo{author}{Wulferding, D.} \emph{et~al.}
\newblock \bibinfo{title}{{Emergent nematicity and intrinsic versus extrinsic
  electronic scattering processes in the kagome metal CsV$_3$Sb$_5$}}.
\newblock \emph{\bibinfo{journal}{Phys. Rev. Res.}}
  \textbf{\bibinfo{volume}{4}}, \bibinfo{pages}{023215} (\bibinfo{year}{2022}).

\bibitem{Song2022}
\bibinfo{author}{Song, D.} \emph{et~al.}
\newblock \bibinfo{title}{{Orbital ordering and fluctuations in a kagome
  superconductor CsV$_3$Sb$_5$}}.
\newblock \emph{\bibinfo{journal}{Science China Physics, Mechanics {\&}
  Astronomy}} \textbf{\bibinfo{volume}{65}}, \bibinfo{pages}{247462}
  (\bibinfo{year}{2022}).

\bibitem{Nie2022}
\bibinfo{author}{Nie, L.} \emph{et~al.}
\newblock \bibinfo{title}{Charge-density-wave-driven electronic nematicity in a
  kagome superconductor}.
\newblock \emph{\bibinfo{journal}{Nature}} \textbf{\bibinfo{volume}{604}},
  \bibinfo{pages}{59--64} (\bibinfo{year}{2022}).

\bibitem{Luo2022_2}
\bibinfo{author}{Luo, J.} \emph{et~al.}
\newblock \bibinfo{title}{{Possible star-of-David pattern charge density wave
  with additional modulation in the kagome superconductor CsV$_3$Sb$_5$}}.
\newblock \emph{\bibinfo{journal}{npj Quantum Materials}}
  \textbf{\bibinfo{volume}{7}}, \bibinfo{pages}{30} (\bibinfo{year}{2022}).

\bibitem{Zhou2022}
\bibinfo{author}{Zhou, S.} \& \bibinfo{author}{Wang, Z.}
\newblock \bibinfo{title}{Chern fermi pocket, topological pair density wave,
  and charge-4e and charge-6e superconductivity in kagom{\'e} superconductors}.
\newblock \emph{\bibinfo{journal}{Nature Communications}}
  \textbf{\bibinfo{volume}{13}}, \bibinfo{pages}{7288} (\bibinfo{year}{2022}).

\bibitem{PhysRevX.13.031030}
\bibinfo{author}{Li, H.} \emph{et~al.}
\newblock \bibinfo{title}{Small fermi pockets intertwined with charge stripes
  and pair density wave order in a kagome superconductor}.
\newblock \emph{\bibinfo{journal}{Phys. Rev. X}} \textbf{\bibinfo{volume}{13}},
  \bibinfo{pages}{031030} (\bibinfo{year}{2023}).

\bibitem{Yin_2021}
\bibinfo{author}{Yin, Q.} \emph{et~al.}
\newblock \bibinfo{title}{{Superconductivity and Normal-State Properties of
  Kagome Metal RbV$_3$Sb$_5$ Single Crystals}}.
\newblock \emph{\bibinfo{journal}{Chinese Physics Letters}}
  \textbf{\bibinfo{volume}{38}}, \bibinfo{pages}{037403}
  (\bibinfo{year}{2021}).
\newblock \urlprefix\url{https://dx.doi.org/10.1088/0256-307X/38/3/037403}.

\bibitem{PhysRevMaterials.5.034801}
\bibinfo{author}{Ortiz, B.~R.} \emph{et~al.}
\newblock \bibinfo{title}{{Superconductivity in the Z2 kagome metal
  KV$_3$Sb$_5$}}.
\newblock \emph{\bibinfo{journal}{Phys. Rev. Mater.}}
  \textbf{\bibinfo{volume}{5}}, \bibinfo{pages}{034801} (\bibinfo{year}{2021}).

\bibitem{Duan2021}
\bibinfo{author}{Duan, W.} \emph{et~al.}
\newblock \bibinfo{title}{{Nodeless superconductivity in the kagome metal
  CsV$_3$Sb$_5$}}.
\newblock \emph{\bibinfo{journal}{Science China Physics, Mechanics {\&}
  Astronomy}} \textbf{\bibinfo{volume}{64}}, \bibinfo{pages}{107462}
  (\bibinfo{year}{2021}).

\bibitem{Ni_2021}
\bibinfo{author}{Ni, S.} \emph{et~al.}
\newblock \bibinfo{title}{{Anisotropic Superconducting Properties of Kagome
  Metal CsV$_3$Sb$_5$}}.
\newblock \emph{\bibinfo{journal}{Chinese Physics Letters}}
  \textbf{\bibinfo{volume}{38}}, \bibinfo{pages}{057403}
  (\bibinfo{year}{2021}).
\newblock \urlprefix\url{https://dx.doi.org/10.1088/0256-307X/38/5/057403}.

\bibitem{Zhong2023}
\bibinfo{author}{Zhong, Y.} \emph{et~al.}
\newblock \bibinfo{title}{{Nodeless electron pairing in CsV$_3$Sb$_5$-derived
  kagome superconductors}}.
\newblock \emph{\bibinfo{journal}{Nature}} \textbf{\bibinfo{volume}{617}},
  \bibinfo{pages}{488--492} (\bibinfo{year}{2023}).

\bibitem{Roppongi2023}
\bibinfo{author}{Roppongi, M.} \emph{et~al.}
\newblock \bibinfo{title}{{Bulk evidence of anisotropic s-wave pairing with no
  sign change in the kagome superconductor CsV$_3$Sb$_5$}}.
\newblock \emph{\bibinfo{journal}{Nature Communications}}
  \textbf{\bibinfo{volume}{14}}, \bibinfo{pages}{667} (\bibinfo{year}{2023}).

\bibitem{Mu_2021}
\bibinfo{author}{Mu, C.} \emph{et~al.}
\newblock \bibinfo{title}{{S-Wave Superconductivity in Kagome Metal
  CsV$_3$Sb$_5$ Revealed by 121/123Sb NQR and 51V NMR Measurements}}.
\newblock \emph{\bibinfo{journal}{Chinese Physics Letters}}
  \textbf{\bibinfo{volume}{38}}, \bibinfo{pages}{077402}
  (\bibinfo{year}{2021}).
\newblock \urlprefix\url{https://dx.doi.org/10.1088/0256-307X/38/7/077402}.

\bibitem{PhysRevB.104.174507}
\bibinfo{author}{Yin, L.} \emph{et~al.}
\newblock \bibinfo{title}{{Strain-sensitive superconductivity in the kagome
  metals KV$_3$Sb$_5$ and CsV$_3$Sb$_5$ probed by point-contact spectroscopy}}.
\newblock \emph{\bibinfo{journal}{Phys. Rev. B}}
  \textbf{\bibinfo{volume}{104}}, \bibinfo{pages}{174507}
  (\bibinfo{year}{2021}).

\bibitem{Gupta2022}
\bibinfo{author}{Gupta, R.} \emph{et~al.}
\newblock \bibinfo{title}{{Microscopic evidence for anisotropic multigap
  superconductivity in the CsV$_3$Sb$_5$ kagome superconductor}}.
\newblock \emph{\bibinfo{journal}{npj Quantum Materials}}
  \textbf{\bibinfo{volume}{7}}, \bibinfo{pages}{49} (\bibinfo{year}{2022}).

\bibitem{PhysRevB.108.144508}
\bibinfo{author}{Holb\ae{}k, S.~C.}, \bibinfo{author}{Christensen, M.~H.},
  \bibinfo{author}{Kreisel, A.} \& \bibinfo{author}{Andersen, B.~M.}
\newblock \bibinfo{title}{Unconventional superconductivity protected from
  disorder on the kagome lattice}.
\newblock \emph{\bibinfo{journal}{Phys. Rev. B}}
  \textbf{\bibinfo{volume}{108}}, \bibinfo{pages}{144508}
  (\bibinfo{year}{2023}).

\bibitem{ge2022discovery}
\bibinfo{author}{Ge, J.} \emph{et~al.}
\newblock \bibinfo{title}{{Discovery of charge-4e and charge-6e
  superconductivity in kagome superconductor CsV$_3$Sb$_5$}}.
\newblock \emph{\bibinfo{journal}{arXiv:2201.10352}}  (\bibinfo{year}{2022}).

\bibitem{zhang2023vortex}
\bibinfo{author}{Zhang, X.} \emph{et~al.}
\newblock \bibinfo{title}{{Vortex phase diagram of kagome superconductor
  CsV$_3$Sb$_5$}} (\bibinfo{year}{2023}).
\newblock \eprint{2306.13297}.

\bibitem{varma2023extended}
\bibinfo{author}{Varma, C.~M.} \& \bibinfo{author}{Wang, Z.}
\newblock \bibinfo{title}{{Extended superconducting fluctuation region and 6e
  and 4e flux-quantization in a Kagome compound with a normal state of
  3Q-order}} (\bibinfo{year}{2023}).
\newblock \eprint{2307.00448}.

\bibitem{PhysRevLett.126.247001}
\bibinfo{author}{Chen, K.~Y.} \emph{et~al.}
\newblock \bibinfo{title}{{Double Superconducting Dome and Triple Enhancement
  of $T_{c}$ in the Kagome Superconductor CsV$_3$Sb$_5$ under High Pressure}}.
\newblock \emph{\bibinfo{journal}{Phys. Rev. Lett.}}
  \textbf{\bibinfo{volume}{126}}, \bibinfo{pages}{247001}
  (\bibinfo{year}{2021}).

\bibitem{PhysRevResearch.3.043018}
\bibinfo{author}{Wang, N.~N.} \emph{et~al.}
\newblock \bibinfo{title}{{Competition between charge-density-wave and
  superconductivity in the kagome metal RbV$_3$Sb$_5$}}.
\newblock \emph{\bibinfo{journal}{Phys. Rev. Res.}}
  \textbf{\bibinfo{volume}{3}}, \bibinfo{pages}{043018} (\bibinfo{year}{2021}).

\bibitem{PhysRevB.103.L220504}
\bibinfo{author}{Du, F.} \emph{et~al.}
\newblock \bibinfo{title}{{Pressure-induced double superconducting domes and
  charge instability in the kagome metal KV$_3$Sb$_5$}}.
\newblock \emph{\bibinfo{journal}{Phys. Rev. B}}
  \textbf{\bibinfo{volume}{103}}, \bibinfo{pages}{L220504}
  (\bibinfo{year}{2021}).

\bibitem{PhysRevMaterials.6.L041801}
\bibinfo{author}{Oey, Y.~M.} \emph{et~al.}
\newblock \bibinfo{title}{{Fermi level tuning and double-dome superconductivity
  in the kagome metal CsV$_3$Sb$_{5-x}$Sn$_x$}}.
\newblock \emph{\bibinfo{journal}{Phys. Rev. Mater.}}
  \textbf{\bibinfo{volume}{6}}, \bibinfo{pages}{L041801}
  (\bibinfo{year}{2022}).

\bibitem{PhysRevMaterials.6.074802}
\bibinfo{author}{Oey, Y.~M.}, \bibinfo{author}{Kaboudvand, F.},
  \bibinfo{author}{Ortiz, B.~R.}, \bibinfo{author}{Seshadri, R.} \&
  \bibinfo{author}{Wilson, S.~D.}
\newblock \bibinfo{title}{{Tuning charge density wave order and
  superconductivity in the kagome metals KV$_3$Sb$_{5-x}$Sn$_x$ and
  RbV$_3$Sb$_{5-x}$Sn$_x$}}.
\newblock \emph{\bibinfo{journal}{Phys. Rev. Mater.}}
  \textbf{\bibinfo{volume}{6}}, \bibinfo{pages}{074802} (\bibinfo{year}{2022}).

\bibitem{Yu2021}
\bibinfo{author}{Yu, F.~H.} \emph{et~al.}
\newblock \bibinfo{title}{Unusual competition of superconductivity and
  charge-density-wave state in a compressed topological kagome metal}.
\newblock \emph{\bibinfo{journal}{Nature Communications}}
  \textbf{\bibinfo{volume}{12}}, \bibinfo{pages}{3645} (\bibinfo{year}{2021}).

\bibitem{PhysRevB.107.174107}
\bibinfo{author}{Tsirlin, A.~A.} \emph{et~al.}
\newblock \bibinfo{title}{{Effect of nonhydrostatic pressure on the
  superconducting kagome metal CsV$_3$Sb$_5$}}.
\newblock \emph{\bibinfo{journal}{Phys. Rev. B}}
  \textbf{\bibinfo{volume}{107}}, \bibinfo{pages}{174107}
  (\bibinfo{year}{2023}).

\bibitem{Feng2023}
\bibinfo{author}{Feng, X.~Y.} \emph{et~al.}
\newblock \bibinfo{title}{{Commensurate-to-incommensurate transition of
  charge-density-wave order and a possible quantum critical point in
  pressurized kagome metal CsV$_3$Sb$_5$}}.
\newblock \emph{\bibinfo{journal}{npj Quantum Materials}}
  \textbf{\bibinfo{volume}{8}}, \bibinfo{pages}{23} (\bibinfo{year}{2023}).

\bibitem{Zheng2022}
\bibinfo{author}{Zheng, L.} \emph{et~al.}
\newblock \bibinfo{title}{{Emergent charge order in pressurized kagome
  superconductor CsV$_3$Sb$_5$}}.
\newblock \emph{\bibinfo{journal}{Nature}} \textbf{\bibinfo{volume}{611}},
  \bibinfo{pages}{682--687} (\bibinfo{year}{2022}).

\bibitem{Sur2023}
\bibinfo{author}{Sur, Y.}, \bibinfo{author}{Kim, K.-T.}, \bibinfo{author}{Kim,
  S.} \& \bibinfo{author}{Kim, K.~H.}
\newblock \bibinfo{title}{{Optimized superconductivity in the vicinity of a
  nematic quantum critical point in the kagome superconductor
  Cs(V$_{1-x}$Ti$_x$)$_3$Sb$_5$}}.
\newblock \emph{\bibinfo{journal}{Nature Communications}}
  \textbf{\bibinfo{volume}{14}}, \bibinfo{pages}{3899} (\bibinfo{year}{2023}).

\bibitem{yang2022titanium}
\bibinfo{author}{Yang, H.} \emph{et~al.}
\newblock \bibinfo{title}{{Titanium doped kagome superconductor
  CsV$_{3-x}$Ti$_x$Sb$_5$ and two distinct phases}}.
\newblock \emph{\bibinfo{journal}{Science Bulletin}}
  \textbf{\bibinfo{volume}{67}}, \bibinfo{pages}{2176--2185}
  (\bibinfo{year}{2022}).

\bibitem{Kautzsch2023}
\bibinfo{author}{Kautzsch, L.} \emph{et~al.}
\newblock \bibinfo{title}{{Incommensurate charge-stripe correlations in the
  kagome superconductor CsV$_3$Sb$_{5-x}$Sn$_x$}}.
\newblock \emph{\bibinfo{journal}{npj Quantum Materials}}
  \textbf{\bibinfo{volume}{8}}, \bibinfo{pages}{37} (\bibinfo{year}{2023}).

\bibitem{Li2022}
\bibinfo{author}{Li, H.} \emph{et~al.}
\newblock \bibinfo{title}{{Discovery of conjoined charge density waves in the
  kagome superconductor CsV$_3$Sb$_5$}}.
\newblock \emph{\bibinfo{journal}{Nature Communications}}
  \textbf{\bibinfo{volume}{13}}, \bibinfo{pages}{6348} (\bibinfo{year}{2022}).

\bibitem{ortiz2023complete}
\bibinfo{author}{Ortiz, B.~R.} \emph{et~al.}
\newblock \bibinfo{title}{{Complete miscibility amongst the AV$_3$Sb$_5$ kagome
  superconductors: Design of mixed A-site AV$_3$Sb$_5$ (A: K, Rb, Cs) alloys}}.
\newblock \emph{\bibinfo{journal}{Physical Review Materials}}
  \textbf{\bibinfo{volume}{7}}, \bibinfo{pages}{014801} (\bibinfo{year}{2023}).

\bibitem{liu2022evolution}
\bibinfo{author}{Liu, M.} \emph{et~al.}
\newblock \bibinfo{title}{{Evolution of superconductivity and charge density
  wave through Ta and Mo doping in CsV$_3$Sb$_5$}}.
\newblock \emph{\bibinfo{journal}{Physical Review B}}
  \textbf{\bibinfo{volume}{106}}, \bibinfo{pages}{L140501}
  (\bibinfo{year}{2022}).

\bibitem{li2022tuning}
\bibinfo{author}{Li, Y.} \emph{et~al.}
\newblock \bibinfo{title}{{Tuning the competition between superconductivity and
  charge order in the kagome superconductor Cs(V$_{1-x}$Nb$_x$)$_3$Sb$_5$}}.
\newblock \emph{\bibinfo{journal}{Physical Review B}}
  \textbf{\bibinfo{volume}{105}}, \bibinfo{pages}{L180507}
  (\bibinfo{year}{2022}).

\bibitem{zhou2023effects}
\bibinfo{author}{Zhou, X.} \emph{et~al.}
\newblock \bibinfo{title}{{Effects of niobium doping on the charge density wave
  and electronic correlations in the kagome metal
  Cs(V$_{1-x}$Nb$_x$)$_3$Sb$_5$}}.
\newblock \emph{\bibinfo{journal}{Physical Review B}}
  \textbf{\bibinfo{volume}{107}}, \bibinfo{pages}{125124}
  (\bibinfo{year}{2023}).

\bibitem{xiao2023evolution}
\bibinfo{author}{Xiao, Q.} \emph{et~al.}
\newblock \bibinfo{title}{{Evolution of charge density waves from
  three-dimensional to quasi-two-dimensional in Kagome superconductors
  Cs(V$_{1-x}$M$_x$)$_3$Sb$_5$ (M = Nb, Ta)}}.
\newblock \emph{\bibinfo{journal}{arXiv:2304.01740}}  (\bibinfo{year}{2023}).

\bibitem{li2022strong}
\bibinfo{author}{Li, J.} \emph{et~al.}
\newblock \bibinfo{title}{{Strong-coupling superconductivity and weak vortex
  pinning in Ta-doped CsV$_3$Sb$_5$ single crystals}}.
\newblock \emph{\bibinfo{journal}{Physical Review B}}
  \textbf{\bibinfo{volume}{106}}, \bibinfo{pages}{214529}
  (\bibinfo{year}{2022}).

\bibitem{liu2023doping}
\bibinfo{author}{Liu, Y.} \emph{et~al.}
\newblock \bibinfo{title}{{Doping evolution of superconductivity, charge order,
  and band topology in hole-doped topological kagome superconductors Cs
  (V$_{1-x}$Ti$_x$)$_3$Sb$_5$}}.
\newblock \emph{\bibinfo{journal}{Physical Review Materials}}
  \textbf{\bibinfo{volume}{7}}, \bibinfo{pages}{064801} (\bibinfo{year}{2023}).

\bibitem{hou2023effect}
\bibinfo{author}{Hou, J.} \emph{et~al.}
\newblock \bibinfo{title}{{Effect of hydrostatic pressure on the unconventional
  charge density wave and superconducting properties in two distinct phases of
  doped kagome superconductors CsV$_{3-x}$Ti$_x$Sb$_5$}}.
\newblock \emph{\bibinfo{journal}{Physical Review B}}
  \textbf{\bibinfo{volume}{107}}, \bibinfo{pages}{144502}
  (\bibinfo{year}{2023}).

\bibitem{ding2022effect}
\bibinfo{author}{Ding, G.}, \bibinfo{author}{Wo, H.}, \bibinfo{author}{Gu, Y.},
  \bibinfo{author}{Gu, Y.} \& \bibinfo{author}{Zhao, J.}
\newblock \bibinfo{title}{{Effect of chromium doping on superconductivity and
  charge density wave order in the kagome metal
  Cs(V$_{1-x}$Cr$_x$)$_3$Sb$_5$}}.
\newblock \emph{\bibinfo{journal}{Physical Review B}}
  \textbf{\bibinfo{volume}{106}}, \bibinfo{pages}{235151}
  (\bibinfo{year}{2022}).

\bibitem{liu2022enhancement}
\bibinfo{author}{Liu, Y.} \emph{et~al.}
\newblock \bibinfo{title}{{Enhancement of superconductivity and suppression of
  charge-density wave in As-doped CsV$_3$Sb$_5$}}.
\newblock \emph{\bibinfo{journal}{Physical Review Materials}}
  \textbf{\bibinfo{volume}{6}}, \bibinfo{pages}{124803} (\bibinfo{year}{2022}).

\bibitem{capa2023electron}
\bibinfo{author}{Capa~Salinas, A.~N.} \emph{et~al.}
\newblock \bibinfo{title}{{Electron-hole asymmetry in the phase diagram of
  carrier-tuned CsV$_3$Sb$_5$}}.
\newblock \emph{\bibinfo{journal}{Frontiers in Electronic Materials}}
  \textbf{\bibinfo{volume}{3}}, \bibinfo{pages}{1257490}
  (\bibinfo{year}{2023}).

\bibitem{oey2022fermi}
\bibinfo{author}{Oey, Y.~M.} \emph{et~al.}
\newblock \bibinfo{title}{{Fermi level tuning and double-dome superconductivity
  in the kagome metal CsV$_3$Sb$_{5-x}$Sn$_x$}}.
\newblock \emph{\bibinfo{journal}{Physical Review Materials}}
  \textbf{\bibinfo{volume}{6}}, \bibinfo{pages}{L041801}
  (\bibinfo{year}{2022}).

\bibitem{lei2023band}
\bibinfo{author}{Lei, X.} \emph{et~al.}
\newblock \bibinfo{title}{{Band splitting and enhanced charge density wave
  modulation in Mn-implanted CsV$_3$Sb$_5$}}.
\newblock \emph{\bibinfo{journal}{Nanoscale Advances}}
  \textbf{\bibinfo{volume}{5}}, \bibinfo{pages}{2785--2793}
  (\bibinfo{year}{2023}).

\bibitem{werhahn2022kagome}
\bibinfo{author}{Werhahn, D.} \emph{et~al.}
\newblock \bibinfo{title}{{The kagom{\'e} metals RbTi$_3$Bi$_5$ and
  CsTi$_3$Bi$_5$}}.
\newblock \emph{\bibinfo{journal}{Zeitschrift f{\"u}r Naturforschung B}}
  \textbf{\bibinfo{volume}{77}}, \bibinfo{pages}{757--764}
  (\bibinfo{year}{2022}).

\bibitem{PhysRevLett.131.026701}
\bibinfo{author}{Liu, B.} \emph{et~al.}
\newblock \bibinfo{title}{{Tunable Van Hove Singularity without Structural
  Instability in Kagome Metal CsTi$_3$Bi$_5$}}.
\newblock \emph{\bibinfo{journal}{Phys. Rev. Lett.}}
  \textbf{\bibinfo{volume}{131}}, \bibinfo{pages}{026701}
  (\bibinfo{year}{2023}).

\bibitem{Wang_2023}
\bibinfo{author}{Wang, Y.} \emph{et~al.}
\newblock \bibinfo{title}{{Flat Band and Z2 Topology of Kagome Metal
  CsTi$_3$Bi$_5$}}.
\newblock \emph{\bibinfo{journal}{Chinese Physics Letters}}
  \textbf{\bibinfo{volume}{40}}, \bibinfo{pages}{037102}
  (\bibinfo{year}{2023}).
\newblock \urlprefix\url{https://dx.doi.org/10.1088/0256-307X/40/3/037102}.

\bibitem{li2023electronic}
\bibinfo{author}{Li, H.} \emph{et~al.}
\newblock \bibinfo{title}{{Electronic nematicity without charge density waves
  in titanium-based kagome metal}}.
\newblock \emph{\bibinfo{journal}{Nature Physics}} \bibinfo{pages}{1--8}
  (\bibinfo{year}{2023}).

\bibitem{yang2022titaniumbased}
\bibinfo{author}{Yang, H.} \emph{et~al.}
\newblock \bibinfo{title}{{Titanium-based kagome superconductor CsTi$_3$Bi$_5$
  and topological states}} (\bibinfo{year}{2022}).
\newblock \eprint{2209.03840}.

\bibitem{liu2023superconductivity}
\bibinfo{author}{Liu, Y.} \emph{et~al.}
\newblock \bibinfo{title}{{Superconductivity emerged from density-wave order in
  a kagome bad metal}}.
\newblock \emph{\bibinfo{journal}{arXiv:2309.13514}}  (\bibinfo{year}{2023}).

\bibitem{PhysRevB.104.235139}
\bibinfo{author}{Pokharel, G.} \emph{et~al.}
\newblock \bibinfo{title}{{Electronic properties of the topological kagome
  metals YV$_6$Sn$_6$ and GdV$_6$Sn$_6$}}.
\newblock \emph{\bibinfo{journal}{Phys. Rev. B}}
  \textbf{\bibinfo{volume}{104}}, \bibinfo{pages}{235139}
  (\bibinfo{year}{2021}).

\bibitem{doi:10.7566/JPSJ.90.124704}
\bibinfo{author}{Ishikawa, H.}, \bibinfo{author}{Yajima, T.},
  \bibinfo{author}{Kawamura, M.}, \bibinfo{author}{Mitamura, H.} \&
  \bibinfo{author}{Kindo, K.}
\newblock \bibinfo{title}{{GdV$_6$Sn$_6$: A Multi-carrier Metal with
  Non-magnetic 3d-electron Kagome Bands and 4f-electron Magnetism}}.
\newblock \emph{\bibinfo{journal}{Journal of the Physical Society of Japan}}
  \textbf{\bibinfo{volume}{90}}, \bibinfo{pages}{124704}
  (\bibinfo{year}{2021}).
\newblock \eprint{https://doi.org/10.7566/JPSJ.90.124704}.

\bibitem{PhysRevB.104.144506}
\bibinfo{author}{Qian, T.} \emph{et~al.}
\newblock \bibinfo{title}{{Revealing the competition between charge density
  wave and superconductivity in CsV$_3$Sb$_5$ through uniaxial strain}}.
\newblock \emph{\bibinfo{journal}{Phys. Rev. B}}
  \textbf{\bibinfo{volume}{104}}, \bibinfo{pages}{144506}
  (\bibinfo{year}{2021}).

\bibitem{guo2023correlated}
\bibinfo{author}{Guo, C.} \emph{et~al.}
\newblock \bibinfo{title}{{Correlated order at the tipping point in the kagome
  metal CsV$_3$Sb$_5$}} (\bibinfo{year}{2023}).
\newblock \eprint{2304.00972}.

\bibitem{xing2023optical}
\bibinfo{author}{Xing, Y.} \emph{et~al.}
\newblock \bibinfo{title}{{Optical Manipulation of the Charge Density Wave
  state in RbV$_3$Sb$_5$}} (\bibinfo{year}{2023}).
\newblock \eprint{2308.04128}.

\bibitem{PhysRevB.106.144504}
\bibinfo{author}{Christensen, M.~H.}, \bibinfo{author}{Birol, T.},
  \bibinfo{author}{Andersen, B.~M.} \& \bibinfo{author}{Fernandes, R.~M.}
\newblock \bibinfo{title}{{Loop currents in AV$_3$Sb$_5$ kagome metals:
  Multipolar and toroidal magnetic orders}}.
\newblock \emph{\bibinfo{journal}{Phys. Rev. B}}
  \textbf{\bibinfo{volume}{106}}, \bibinfo{pages}{144504}
  (\bibinfo{year}{2022}).

\end{thebibliography}

\end{document}